\documentclass[12pt]{article}
\usepackage{a4,epsfig}
\usepackage{times,amsmath,amssymb,latexsym}
\usepackage{rotating}
\usepackage{isolatin1}
\usepackage[cp850]{inputenc}
\usepackage{stmaryrd}

\newcommand{\delbar}{\bar{\partial}}
\newcommand{\del}{\partial}
\newcommand{\no}[1]{{:\!#1\!:}}
\newcommand{\df}{\Rightarrow}

\newcommand{\eqn}[1]{\begin{eqnarray} #1 \end{eqnarray}}
\newcommand{\abs}[1]{\left| #1 \right|}
\newcommand{\ci}{{\rm i}}
\newcommand{\nstress}[1]{{\em #1}}
\newcommand{\pstress}[1]{{\sc #1}}

\newcommand{\cstar}{\widehat{\mathbb{C}}}
\newcommand{\equals}{&\!\!\!=\!\!\!&}

\newcommand{\ds}{\displaystyle}
\newcommand{\phan}{\phantom{\frac{17}{10},\,}}
\def\switchtitlefootnote{\renewcommand{\thefootnote}{\fnsymbol{footnote}}}
\def\switchtextfootnote{\renewcommand{\thefootnote}{\arabic{footnote}}}

\def\abstracttext{
We propose to describe bulk wave functions of fractional quantum
\pstress{Hall} states
in terms of correlators of non-unitary \nstress{$b$/$c$-spin} systems.
These yield a promising conformal field theory analogon of
the composite fermion picture of \pstress{Jain}.
Fractional statistics is described by twist fields which
naturally appear in the \nstress{$b$/$c$-spin} systems. We provide
a geometrical interpretation of our approach in which bulk wave functions
are seen as holomorphic functions over a ramified covering of the
complex plane, where the ramification precisely resembles the fractional
statistics of the quasi-particle excitations in terms of branch points on
the complex plane. To extend \pstress{Jain}'s
main series, we use the concept of composite fermions pairing to spin singlets,
which enjoys a natural description in terms of the particular ${\rm c}=-2$
\nstress{$b$/$c$-spin} system as known from the \pstress{Haldane-Rezayi}
state. In this way we derive conformal field theory
proposals for lowest \pstress{Landau} level bulk wave
functions for more general filling fractions. We obtain a natural
classification of the experimentally confirmed filling fractions which does
not contain prominent unobserved fillings. Furthermore, our scheme fits
together with classifications in terms of $K$-matrices of effective
multilayer theories leading to striking restrictions of these
coupling matrices.
}

\begin{document}
%-----------------------------------------------------------
% Title page
%-----------------------------------------------------------
\switchtitlefootnote \pagestyle{empty}
\begin{titlepage}

\begin{center}

\vspace*{-1.0cm} {\hbox to \hsize{\hfill cond-mat/0208429}} {\hbox to
\hsize{\hfill ITP-UH-20/02}} \vspace*{1.0cm}

\vspace*{0.5cm} {\Large \bf
        A Novel CFT Approach to Bulk Wave Functions\\[1ex]
    in the Fractional Quantum Hall Effect}

\vspace*{1cm} {Michael Flohr\footnote[2]{\parbox[t]{12cm}{\tt
        Michael.Flohr@ITP.Uni-Hannover.De}}} and
    Klaus Osterloh\footnote[1]{\parbox[t]{12cm}{\tt
        Klaus.Osterloh@ITP.Uni-Hannover.De}}\\[1cm]

{\em
        Institut f{\"u}r Theoretische Physik\\
        Universit{\"a}t Hannover\\
        Appelstra{\ss}e 2, 30167 Hannover, Germany}

\vspace*{1cm} {\small August 22, 2002}

\vspace*{2cm} {\small \bf Abstract}\\[0.5cm]

\parbox{13.5cm}
{\small \abstracttext }
\end{center}
\end{titlepage}

\newpage
\setcounter{page}{1} \switchtextfootnote \pagestyle{plain}
%
% Section - Introduction
%
\section{Introduction}
The fractional quantum \pstress{Hall} effect (FQHE) is one of the most
fascinating and striking phenomena in condensed matter physics
\cite{Tsui:1982yy}. Certain numbers, the filling fractions $\nu\in\mathbb{Q}$,
can be observed with an extremely high precision in terms of the \pstress{Hall}
conductivity $\sigma_{{\rm H}}=\nu$ in natural units. These numbers are
independent of many physical details such as the geometry of the sample, its
purity, the temperature -- at least within large bounds. The enigmatic and
fascinating aspect of this phenomenon is that only a certain set of these
fractional numbers $\nu$ can be observed in experiments: despite ongoing
attempts in varying the purity (or disorder), the external magnetic field and
various other parameters, the set of observed fractions has not much changed
over the last few years
\cite{Stormer:1999aa,Stormer:2000aa,Stormer:2001aa,Stormer:2002aa}.

It was realized quite early that the FQHE shows all signs of universality and
large scale behavior  \cite{Frohlich:xz, Frohlich:wb}. Independence of the
geometrical details of the probe and its size hints towards an effective purely
topological field theory description. Indeed, since the quantum \pstress{Hall}
effect is essentially a (2$+$1)-dimensional problem, the effective theory is
regarded to be dominated by the topological \pstress{Chern-Simons} term
$a\wedge {\rm d}a$ instead of the \pstress{Maxwell} term ${{\rm tr}}F^2$ Some
good reviews on the theory of the FQHE are
\cite{Halperin:1983aa,Prange:1987aa,Stone:ft,Wu:1990ee}.

However, one is ultimately interested in a microscopic description of the FQHE.
One may start with the task of finding eigenstates of an exact microscopic
\pstress{Hamiltonian}. This can be done numerically for small numbers of
electrons. The great achievement of \pstress{Laughlin} was to realize how a
many-particle wave function looks like if it is to respect a few common sense
symmetry constraints \cite{Laughlin:1983fy}:
\begin{equation}
  \Psi_{\rm Laughlin}(z_1,\,\ldots,\,z_N)
  =\prod_{1\leq i<j\leq N}(z_i - z_j)^{2p+1}\exp\left(
  -\frac{1}{4}\sum_{1\leq i\leq N}|z_i|^2\right)\,.
\end{equation}
We now know that \pstress{Laughlin}'s wave functions are extremely good
approximations to the true ground states, and they are exact solutions for
\pstress{Hamiltonians} with certain short-range electron-electron interactions.
They describe fractional quantum \pstress{Hall} states (FQH states) with
filling $\nu=1/(2p+1)$, $p\in\mathbb{Z_+}$. Soon after, various so-called
hierarchical schemes were developed yielding ground state wave functions for
other rational filling factors
\cite{Halperin:fn,Halperin:1992mh,Haldane:1983xm,Jain:1989tx, Read:1990xx}. The
important point to note here is that the ground state eigenfunctions are
time-independent up to a trivial global phase. Thus, one might view them as
solutions of a (2$+$0)-dimensional problem. This is, more or less, the main
idea behind all attempts to describe the bulk wave functions in terms of
conformal field theory (CFT) correlators.

The \pstress{Laughlin} wave functions describe special incompressible quantum
states of the electrons, so-called quantum droplets. Incompressibility is
connected to the existence of \nstress{energy gapless} excitations on the
border of the quantum state \cite{Halperin:1981ug, Frohlich:xz, Frohlich:wb,
Wen:1992vi, Blok:1990mc,Milovanovic:1996nj,
 Cappelli:1992kf, Cappelli:1996pg}. The latter can successfully be described in terms of CFTs with current
algebras as chiral symmetries. Furthermore, there is an exact equivalence
between the (2$+$1)-dimensional \pstress{Chern-Simons} theory in the bulk and
the (1$+$1)-dimensional conformal field theory on the boundary describing the
edge excitations \cite{Witten:1988hf}. We note here that, naturally, such CFTs
have to be unitary, since they describe the time evolution of spatially
one-dimensional waves propagating on $S^1$.

However, \pstress{Laughlin}'s bulk wave functions in a static
(2$+$0)-dimensional setting show a striking resemblance to correlation
functions of a free \pstress{Euclidean} CFT put on the (compactified) complex
plane. This resemblance has motivated quite a number of works trying to find a
CFT description of bulk wave functions in the FQHE, e.g.\ \cite{Moore:ks,
Cristofano:fg, Fubini:1990bw, Flohr:1996aa}. Most approaches assumed from the
beginning that these ``bulk'' theories are unitary. We stress here that this
assumption is void, since the bulk wave functions one typically wants to
represent are time-independent eigenfunctions. Moreover, most approaches
represented the bulk wave functions in terms of building blocks belonging to
classes of CFTs with continuous parameters, e.g., the \pstress{Gaussian} ${\rm
c}=1$ CFTs. The immanent problem with these approaches is that there is no
principle selecting the wave functions for experimentally observed filling
fractions. Therefore, almost all approaches so far easily accommodate arbitrary
rational filling factors. On the other hand, it is not entirely surprising that
the bulk wave function should have something to do with CFT. As mentioned
above, the observable quantities of the quantum \pstress{Hall} system are
largely independent of the precise form and size of the sample. Thus, the
normalized charge distributions of the electrons should be invariant under
scaling (up to an exponential factor) and area preserving changes of the shape
of the sample. The first symmetry is linked to conformal invariance, the latter
to the ${\cal W}_{1+\infty}$-algebra \cite{Cappelli:1992yv, Flohr:1993ga}. In
fact, it is known that in the two-dimensional case global scaling invariance
implies full conformal invariance under certain benign circumstances.

Interestingly, there exists a particularly enigmatic FQH state, i.e.~the
\pstress{Haldane-Rezayi} state with $\nu=5/2$. This is one of the very few
states with an even denominator filling. Of course, attempts have been made to
describe proposed bulk wave functions for this state with the help of CFT
correlators, see e.g.\ \cite{Wen:ha, Milovanovic:1996nj, Moore:ks,
Flohr:1997aa}. In this case, however, it turned out that this can only be done
if the CFT in question has central charge ${\rm c}=-2$. Thus, for this FQH
state we necessarily have to use a non-unitary theory. On the other hand, this
CFT is well known, it is the \nstress{$b$/$c$-spin} system of two
anti-commuting fields with spins one and zero, respectively. Therefore, it
naturally yields precisely the object one had expected in this FQH state,
namely spin singlet states of paired electrons. In addition, the ${\rm c}=-2$
CFT contains a $\mathbb{Z}_2$-twisted sector created by a primary field $\mu$
of conformal scaling dimension $h_{\mu}=-1/8$, which accurately describes the
effect of single flux quanta piercing the quantum droplet. Thus, this theory
successfully characterizes the ground state and its physically expected
excitations with the correct fractional statistics, and only these.

The present paper takes the success of the bulk wave description of the
\pstress{Haldane-Rezayi} FQH state via a non-unitary spin-system CFT
as a starting point to revisit the question, how FQH state bulk wave functions
can be represented in terms of CFTs. In contrast to other
approaches we will drop the assumption that these CFTs
should be unitary because there is no physical reason for it. This enables us
to concentrate on a different class of CFTs, namely the
\nstress{$b$/$c$-spin} systems of two anti-commuting fields of spins $j$ and
$(1-j)$, respectively. Locality forces $j\in\mathbb{Z}/2$ such that we
confine ourselves to a discrete series of CFTs. It will
turn out that our ansatz not only naturally
explains all experimentally observed
filling fractions, but, in addition, does not predict new unobserved series.

Besides these convenient features our approach yields a beautiful geometrical
picture for the CFTs we use to represent the bulk
wave functions. Additionally, we find correlations of spin $j$ (or
spin $1-j$) composite fermions with flux quanta of precisely the fractional
statistics which are theoretically predicted from first principles. These
statistics, say $1/m$, manifest themselves naturally in the presence of
$\mathbb{Z}_m$-twists which in turn have the geometrical meaning of replacing
the complex plane by an $m$-fold ramified covering of itself. Thus, the bulk
wave functions finally are recast in a language of complex analysis, i.e.,
$j$- or $(1-j)$-differentials on $\mathbb{Z}_m$-symmetric
\pstress{Riemann} surfaces.

Most of the observed filling fractions $\nu\in\mathbb{Q}$ have an odd
denominator, which comes from the basic fact that the elementary entities in
the quantum \pstress{Hall} system are fermions. It turns out that unpaired
fermions correspond to spin systems with spin $j$ half-integer (remember that
the paired electrons singlets in the \pstress{Haldane-Rezayi} state were
described by an integer-spin system). An essential part of our paper is that we
will propose a new hierarchical scheme in which filling fractions can be
derived from others by means of forming more and more paired singlets. Besides
\pstress{Jain}'s principle series, this yields further series precisely
catching all confirmed filling fractions. Unobserved filling fractions are no
problem within our scheme, since they all lie at the far end of our series or
are characterized by series of higher order. In contrast to this, most other
hierarchical schemes predict certain unobserved fractions, since prominent
experimentally confirmed ones can only be realized at a certain order $k$
within the hierarchy while others obtained at smaller orders of the hierarchy
do not show up in experiments. The problem is the lack of a physical reason why
the corresponding low order FQH state does not exist, but the higher order FQH
state derived from it in the hierarchy. Thus, we believe that our scheme
provides a natural explanation for the completeness of the set of
experimentally accessible filling fractions which does not run into this
problem.

Our paper proceeds as follows: To be as self-contained as possible we collect
the essential formulae and concepts of CFT in section two. We are not very
general here, since we only concentrate on those facts which are relevant for
the special CFTs, i.e.~the spin systems, that we will use throughout the
reminder of the paper. The reader unfamiliar with CFT might consult
\cite{Itzykson:bk, DiFrancesco:1997aa}

In section three, we briefly review the basic idea of \pstress{Laughlin}
leading to his seminal trial wave functions. Furthermore, we present the
appropriate generalization of these within the
picture of \pstress{Jain} which allows to describe a large class of FQH states
in terms of an effective integer quantum \pstress{Hall} effect (IQHE) of
effective elementary particles, the composite fermions (CF).
We favor this idea,
since our CFT ansatz contains fields which can naturally be
identified with such composite fermions. Moreover, \pstress{Jain}'s picture has
the advantage to realize most of the prominently observed filling fractions
within the first level of its hierarchical scheme.

Section four is the core of our paper. More general
\pstress{Laughlin} type trial wave functions in the lowest
\pstress{Landau} level (LLL)
projection are obtained from multilayer states. In this scheme,
all essential information is encoded in a certain matrix $K$ describing the
coupling of the layers, i.e., of different quantum fluids. It respects
many general principles, such as topological order. Evident physical properties
lead to severe constraints on these $K$-matrices, which will be seen to
coincide nicely with the constraints we find for our spin system CFTs.
Step by step we develop the \nstress{$b$/$c$-spin} approach in terms of
beginning with the simplest case of the
principal main series of \pstress{Jain}'s hierarchy. All other confirmed
filling fractions
are consecutively obtained by pairing of
CFs to spin singlet states. This can be done to a lower or
higher degree resulting in our novel hierarchical scheme.
By this, we do not have to make use of the principle of particle-hole duality
that is not well confirmed by experiment.
Furthermore, our pairing scheme, which is represented
by tensoring the spin CFTs with
additional spin-singlet \nstress{$b$/$c$-spin} systems of central charge
${\rm c}=-2$, puts severe constraints on the possible form of the $K$-matrix,
restricting it essentially to block form. After developing our approach
to the point that all observed filling fractions are obtained and certain
prominent rational numbers, which were never experimentally confirmed, are
ruled out within our approach, we finally provide some predictions for future
experiments.

In the concluding fifth section, we summarize our results and try to put
them into context. We also mention unsolved problems and some directions
for possible research in the future.
The appendix contains some sketchy remarks,
that the space of states of our non-unitary theories appropriately
coincides with the space of states of the (1$+$1)-dimensional theories
describing the edge excitations.
%
% Section - Conformal Field Theory
%
\section{Conformal Field Theory}
During the last decades conformal field theory (CFT) became one of the most
powerful tools of modern theoretical physics \cite{Belavin:1984vu}. Surely, one
of the most important impulses came from statistical mechanics: CFT is
well-known for its applicability to statistical systems at criticality. At a
continuous phase transition the correlation length diverges and the system
becomes scale invariant. In two dimensions this usually implies conformal
invariance of the system. If the corresponding CFT is identified and found to
be rational we can derive the partition function and the problem is solved in a
very elegant and effective way. Apart from that there are lots of phenomena,
for instance bosonization, in solid state physics involving CFT even if it is
not apparent at first sight. Often when geometrical or topological aspects
arise CFT is close at hand and allows to derive global properties without
detailed knowledge of microscopic structures.

We want to stress that this paper deals with bulk CFTs in 2+0 dimensions.
Therefore, it does not make sense to argue about unitarity, time evolution and
similar aspects. Of course, the corresponding edge theory has to be unitary,
but this implies no crucial restriction for the bulk part.

The theories used in our picture are the \nstress{$b$/$c$-spin} systems that
were analyzed in detail by \pstress{Knizhnik} \cite{Knizhnik:kf,Knizhnik:xp}.
They are described by the action
\eqn{S=\int {\rm d}^2z\
b(z)\delbar c(z) + h.c.} Here, $b(z)$ and $c(z)$ are anti-commuting conformal
fields of weight $j\in\mathbb{Z}/2$, and $1-j$ respectively, where $z$ is a
coordinate in the complex plane. Mathematically spoken the fields $b(z)$ and
$c(z)$ describe $j$- and $1-j$-differentials. Therefore, they are directly
related to the cohomology of the topological space they live on. Furthermore,
these theories are chiral CFTs so we can treat the holomorphic part
independently. This nicely coincides with the fact that FQH states considered
in the \nstress{lowest \pstress{Landau} level} (LLL) are described by
holomorphic wave functions.\\ By variation of the action via path integral we
get the equations of motion:
\eqn{(\delbar c(z))b(z')=(\delbar b(z))c(z')=\delta^2(z-z',\
\bar{z}-\bar{z}')\quad ,\quad\delbar b(z)=\delbar c(z)=0\:\: .} In classical
terms we would expect
\eqn{(\delbar c(z))b(z')=(\delbar b(z))c(z')=0\:\:
.\label{classeom}} Thus, the normal-ordered product of the two fields in order
to satisfy \eqref{classeom} reads:
\eqn{\no{b(z)c(z')}=b(z)c(z') - \frac{1}{z-z'}\:\: .\label{nopbc}}
In 2d CFT a product of local chiral operators can be expanded in an operator
valued \pstress{Laurent} series with meromorphic functions as coefficients. In
the evaluation of correlators these so-called \nstress{operator product
expansions} (OPEs) play an important role. The OPEs of the two fields $b(z)$
and $c(z')$ can be read off directly from \eqref{nopbc}:
\eqn{b(z)c(z') \sim \frac{1}{z-z'}\quad ,\quad c(z)b(z') \sim
\frac{1}{z-z'}\:\: .} Here, `$\sim$' denotes `equivalent up to regular terms'.
These regular terms vanish if evaluated in a correlator.

The energy-momentum tensor $T(z)$ of the theory can be derived by varying the
action $S$ with respect to the induced metric. This yields
\eqn{T(z)=(1-j)\no{(\del b(z))c(z)}-j\no{b(z)(\del c(z))}\:\: .}
In principle there are just a few facts we have to know about a general CFT:
the central charge $\rm c$ and the set of conformal weights $\{h_i\}$ of its
fields are two of them. They can be derived by OPEs involving the
energy-momentum tensor:
\eqn{T(z)b(w) &\!\!\!\sim\!\!\!&\frac{j}{(z-w)^2}b(w)+
                                \frac{1}{z-w}\del_wb(w)\:\: ,\label{defeq}\\
     T(z)c(w) &\!\!\!\sim\!\!\!&\frac{1-j}{(z-w)^2}c(w)+
                                \frac{1}{z-w}\del_wc(w)\:\: ,\label{defeqtwo}\\
     T(z)T(w) &\!\!\!\sim\!\!\!&\frac{\frac{1}{2}(-12j^2+12j-2))}{(z-w)^4}+
                         \frac{2}{(z-w)^2}T(w)+\frac{1}{z-w}\del_wT(w)\:\: .
             \label{centralchargebc}}
Equations \eqref{defeq} and \eqref{defeqtwo} can be understood as the
definition of a primary conformal field, the numerator of the first term of the
OPE yields its conformal weight $h$. The third OPE contains a so-called
anomalous term that is not proportional to the field itself or its derivatives.
This term is due to the existence of a central extension of the algebra of
conformal symmetries. In fact, in all CFTs the OPE of $T(z)$ with itself reads
\eqn{T(z)T(w) &\!\!\!\sim\!\!\!&\frac{{\rm c}/2}{(z-w)^4}+
                         \frac{2}{(z-w)^2}T(w)+\frac{1}{z-w}\del_wT(w)\:\: .}
We find \eqn{{\rm c}_{b{\rm /}c{\rm -spin}}=-2(6j^2-6j+1)\:\: .\label{cbc}} For
$j\neq\frac{1}{2}$ the central charge is negative (as $j\in\mathbb{Z}/2$).
Therefore, the \nstress{$b$/$c$-spin} systems used in our scheme
are non-unitary. We will briefly discuss this issue and how our
approach fits together with the unitary edge theories in the appendix.
Furthermore,
there exists an additional symmetry of the action. Under the simultaneous
transformation
\eqn{b(z)\rightarrow b(z)\exp(\ci\alpha)\quad\text{and\quad}
     c(z)\rightarrow c(z)\exp(-\ci\alpha)}
the action remains unchanged. The corresponding conserved spin current $j(z)$
reads:
\eqn{j(z)=-\no{b(z)c(z)}\label{spincurrent}}
with its conserved charge
\eqn{Q_{(\ci\alpha),j}=\frac{1}{2\pi\ci}\oint_0 dz\ (\ci\alpha)j(z)\quad .}
To stress it again, the \nstress{$b$/$c$-spin} systems are directly related to
the topology they live on. In our picture we are interested in
\pstress{Riemann} surfaces (RS) with global $\mathbb{Z}_n$-symmetry. This means
that every branch point is of order $n$ and that all monodromy matrices can be
diagonalized simultaneously . It is sufficient to do the calculation locally
for a single branch point at $z_0$. The results can be directly extended to $m$
branch points.

A $\mathbb{Z}_n$-symmetric RS can be locally represented by a branched covering
of the compactified complex plane
$\left(\widehat{\mathbb{C}}=\mathbb{C}\cup\{\infty\}\right)$ with the following
map:
\eqn{z: \text{RS} \rightarrow \cstar\quad ,\quad z(y)=z_0+y^n
     \quad . \label{rsmap}}
We identify the RS locally by $n$ sheets of $\cstar$ via the inverse map of
\eqref{rsmap}. The \nstress{$b$/$c$-spin} fields living on the RS are therefore
represented by an $n$-dimensional vector of identical copies of the
{$b$/$c$-fields} $b^{(l)}(z)$ and $c^{(l)}(z)$ on the complex plane with
boundary conditions \eqn{\hat{\Pi}_{z_0} b^{(l)}(z)=b^{(l+1)}(z)\:\: ,\quad
l=0,\dots,n-1\:\: ,\quad b^{(n)}(z)=b^{(0)}(z)\:\: ,} where
\eqn{\hat{\Pi}_{z_0}: (z-z_0) \rightarrow (z-z_0)\exp(2\pi\ci)\quad .}
For further investigation we introduce a \pstress{Fourier} basis
\eqn{b_k(z)=\sum_{l=0}^{n-1}\exp\left(\frac{-2\pi\ci(k+j(1-n))l}{n}\right)
     b^{(l)}(z)\:\: ,\nonumber\\
     c_k(z)=\sum_{l=0}^{n-1}\exp\left(\frac{+2\pi\ci(k+j(1-n))l}{n}\right)
     c^{(l)}(z)\:\: .}
This basis diagonalizes $\hat{\Pi}_{z_0}$:
\eqn{\hat{\Pi}_{z_0}b_k(z)=\exp\left(\frac{+2\pi\ci(k+j(1-n))}{n}\right)
     b_k(z)\:\: ,\nonumber\\
     \hat{\Pi}_{z_0}c_k(z)=\exp\left(\frac{-2\pi\ci(k+j(1-n))}{n}\right)
     c_k(z)\:\: .}
As a consequence the conserved spin current \eqref{spincurrent}
becomes single-valued. Therefore, the corresponding charge vector $\alpha_k$
identified with the branch point $z_0$ is
\eqn{\alpha_k=-\frac{k+j(1-n)}{n}\:\: .\label{charge}}
Now we bosonize the theory. This means that we express the spin fields in terms
of exponentials of analytic scalar bosonic fields $\Phi_k$
\eqn{b_k(z)\equals\no{\exp\left(+\ci\Phi_k(z)\right)}\:\: ,\nonumber\\
     c_k(z)\equals\no{\exp\left(-\ci\Phi_k(z)\right)}
     \:\: .\label{bcbosonized}}
This yields that the branch point of the $\mathbb{Z}_n$-symmetric RS is
related to a primary conformal field of the \nstress{$b$/$c$-spin} system:
\eqn{V_{\vec{\alpha}}(z_0)=
     \no{\exp\bigg(\ci\sum_{k=0}^{n-1}\alpha_k\Phi_k\bigg)}\:\: .}
It is called \nstress{Vertex operator} and has conformal weight
\eqn{h_{\vec{\alpha}}=\sum_{k=0}^{n-1}h_{\alpha_k}=
     \sum_{k=0}^{n-1}\left(\frac{1}{2}\alpha_k^2-(j-\frac{1}{2})\alpha_k\right)
     \:\: .\label{vertexweight}}
As these fields are of central importance in our context, let us look a bit
more carefully at them. They are primary conformal fields arising in the most
general case from the CFT of the free boson with an embedded background charge
and are defined by
\eqn{V_\ell(z)=\no{\exp\left(\ci\ell\Phi(z)\right)}\:\: .}
Here, $\Phi(z)$ is a free bosonic field with conformal weight $h=0$. The OPE
reads
\eqn{\Phi(z)\Phi(w)\sim -\ln(z-w)\:\: .\label{freeboson}}
The energy-momentum tensor is given by
\eqn{T(z)=\frac{1}{2}\no{\del_z\Phi(z)\del_z\Phi(z)}+\ci\alpha_0\del^2\Phi(z)
     \quad ,}
where $\alpha_0$ is a background charge placed at infinity. This leads to the
following OPEs:
\eqn{&&T(z)T(w)\sim\frac{\frac{1}{2}(1-12\alpha_0^2)}{(z-w)^4}+
     \frac{2}{(z-w)^2}T(w)+\frac{1}{z-w}\del_wT(w)
     \:\: ,\label{centralcharge}\\
     &&T(z)\del\Phi(w)\sim\frac{2\ci\alpha_0}{(z-w)^3}
       +\frac{1}{(z-w)^2}\del_w\Phi(w)
       +\frac{1}{z-w}\del^2_w\Phi(w)\:\: .\label{delphi}
     }
The background charge is derived from \eqref{centralcharge} and \eqref{cbc}
\eqn{\alpha_0=j-\frac{1}{2}}
in order to bosonize the theory correctly. Furthermore, it follows from
\eqref{delphi} that $\del\Phi$ is not a primary conformal field unless the
background charge vanishes. In fact, even $\Phi$ itself is not primary, as it
is expected from \eqref{freeboson}. Due to the logarithmic term in its OPE the
\nstress{Vertex operator} is the remaining candidate for a primary field. In
fact,
\eqn{T(z)V_\ell(w)&\!\!\!\sim\!\!\!&\sum_{l=0}^{\infty}\frac{(\ci\ell)^l}{l!}
     \left(-\frac{1}{2}\no{\del_z\Phi(z)\del_z\Phi(z)}
     +\ci\alpha_0\del_z^2\Phi\right)\,\no{\Phi(w)^l}\nonumber\\
     &\!\!\!\sim\!\!\!&-\frac{1}{2}\sum_{l=0}^{\infty}\frac{(\ci\ell)^l}{(l-2)!}
     \frac{\no{\Phi(w)^{l-2}}}{(z-w)^2}+
     \sum_{l=0}^{\infty}\frac{(\ci\ell)^l}{(l-1)!}
     \frac{\no{\del_w\Phi(w)\Phi(w)^{l-1}}}{(z-w)}\nonumber\\
     &\!\!\!\phantom{\sim}\!\!\!&+\ci\alpha_0\sum_{l=0}^{\infty}
     \frac{(\ci\ell)^l}{(l-1)!}\frac{\no{\Phi(w)^{l-1}}}{(z-w)^2}\nonumber\\
     &\!\!\!\sim\!\!\!&\frac{\ell^2/2-\alpha_0\ell}{(z-w)^2}\, V_\ell(w)
     +\frac{1}{z-w}\del_wV_\ell(w)}
proves that $V_k(z)$ is primary with conformal weight $h=\ell^2/2-(j-1/2)\ell$.
This is indeed the result of \eqref{vertexweight}.

Having a closer look at \eqref{charge} we immediately find that the charge
vector of the \nstress{Vertex operator} is dominated by the
$\mathbb{Z}_n$-symmetry of the RS. The spin $j$ simply provides an offset which
is just visible in the conformal weight of the fields since the phase is
determined by $\alpha_k\:{\rm mod}\:1$. In addition we have to distinguish
between two different types of fields. First, there are \nstress{twist fields}
that contain the full information of the branch point. Therefore, the charge
vector $\vec{\alpha}$ has to keep track of analytic continuation. For example,
given a $\mathbb{Z}_3$-symmetric RS and $j=3/2$, the charge vector is derived
as
\eqn{\vec{\alpha}_{n=3, j=3/2}=\left(\:1\:,\:2/3\:,\:1/3\:\right)\:\: .}
Secondly, there are \nstress{projective fields}. Their charge components are
identical as if we simply had an $n$-fold copy of $\cstar$. This yields
charge vectors $\vec{\alpha}^{\rm p}$ with
\eqn{\vec{\alpha}^{\rm p}_1=\ldots=
\vec{\alpha}^{\rm p}_n\in\left\{0,\:\frac{1}{n},\:\ldots\:,\:1\right\}
     \:\: .\label{chargev}}
We stress once more the important role of the charge vectors $\vec{\alpha}$.
Besides local chiral fields, whose charge vectors have integer valued
components only, we include fractional ones \eqref{chargev}. The effect of the
corresponding vertex operators is to precisely simulate the action of a branch
point of ramification number $n$. This is exactly the effect we expect from
fractional statistics of quasi-particles. Thus, we incorporate the statistics
into a geometrical setting, where the complex plane is replaced by an $n$-fold
ramified covering of itself, created by flux quanta piercing it.

Naturally, we expect to find the \nstress{projective fields} in order to
describe FQH states in the \nstress{lowest \pstress{Landau} level} (LLL)
projection correctly. Since the bosons $\Phi_k$ are free fields, the
correlators of their \nstress{Vertex operators} read
\eqn{\left\langle\,\Omega\left| V_{\vec{\alpha}_1}(z_1)\cdot\ldots\cdot
     V_{\vec{\alpha}_n}(z_n)\right|0\,\right\rangle=
     \prod_{i<j}^{n}(z_i-z_j)^{\vec{\alpha}_i\cdot\vec{\alpha}_j}
     \:\: ,\label{corr}}
where $\left\langle\,\Omega\,\right|$ is an out-state connected to the
background charge at infinity.

The set of equations \eqref{chargev} and \eqref{corr} including their geometric
features is all we need to derive the LLL projected FQH bulk wave functions for
filling fractions $0\leq\nu\leq 1$.
%
% Section - Laughlin States
%
\section{Laughlin States}
To begin the analysis of the FQHE that was first discovered by \pstress{Tsui},
\pstress{St\"{o}rmer} and \pstress{Gossard} \cite{Tsui:1982yy} it is natural to
start with the \pstress{Laughlin} states. Their wave functions are given by
\eqn{\Psi_{\rm Laughlin}(z_1, \ldots,
z_n)={\cal{N}}\prod_{k<l}^n(z_k-z_l)^{2p+1}\exp\Big(
     -\frac{1}{4}\sum_i^n \abs{z_i}^2 \Big)\:\: ,\label{psilaugh}}
where $p\in\mathbb{N}$, $z_i=x_i+\ci y_i$ is the position of the $i$-th
electron in unified complex coordinates and $\cal{N}$ is a normalization
factor.

These wave functions describe a uniform incompressible quantum fluid of
electrons in the LLL widely separated from each other that obey phase
correlations as if carrying $2p$ flux quanta of the magnetic field. They are
completely anti-symmetric, correspond to filling fractions $\nu=\frac{1}{2p+1}$
and were conceived by \pstress{Laughlin} \cite{Laughlin:1983fy} as the
variational ground state wave functions for the model \pstress{Hamiltonian}
\eqn{{\cal H}=\sum_k^n\left[ \frac{1}{2m}\left(\frac{\hbar}{\rm i}
       \nabla_k-\frac{\rm e}{c}\vec{A}(\vec{r}_k)\right)^2
       +V_{\rm bg}(\vec{r}_k)\right]
       +\sum_{k<l}^n\frac{{\rm e}^2}{\abs{\vec{r}_k-\vec{r}_l}}\:\: .
    }
Here, $V_{\rm bg}$ is a potential of a background charge distribution that
neutralizes the electrons' \pstress{Coulomb} repulsion. This guarantees that
the system is stable. The vector potential is taken in the symmetric gauge
\eqn{\vec{A}(\vec{r})=\frac{B}{2}(-y, x, 0)\quad .}
We stress that electron-electron interaction is a crucial necessity for the
FQHE. In contrast to the IQHE, a one-particle effect involving disorder, the
fractional regime is found to be a strongly correlated system (SCS).

Furthermore, the modulus squared of the wave function is equivalent to the
\pstress{Boltzmann} distribution of a 2d one-component plasma. This yields
further information with respect to the thermodynamic limit,
\eqn{ \abs{\Psi}^2=\exp(-\beta\Phi)\:\: ,} where $\beta=\frac{1}{2p+1}$ and
\eqn{\Phi=-2(2p+1)^2\sum_{k<l}^n\ln\abs{z_k-z_l}
     +\frac{2p+1}{2}\sum_k^n\abs{z_k}^2\:\: .\label{plasma}}
Hence, for small $p$ the system is a liquid rather than a \pstress{Wigner}
crystal.

Another important property is the incompressibility of the \pstress{Laughlin}
states. This leads to the existence of plateaus in the \pstress{Hall}
conductance. The \pstress{Laughlin} ground state can be extended with respect
to quasi-hole excitations by introducing a simple polynomial factor
\eqn{\Psi_{\rm exc.}={\cal{N}}(\zeta_i)\prod_{k,l}(z_k-\zeta_l)
     \prod_{r<s}(z_r-z_s)^{2p+1}
     \exp\Big(-\frac{1}{4}\sum_i\abs{z_i}^2 \Big)\:\: .}
Here, the $\zeta_i$ denote the positions of the quasi-hole excitations. With
respect to \eqref{plasma} the excited states, in contrast to the ground states,
have a non-uniform charge distribution. In comparison with the 2d plasma one
can calculate a charge deficit of $\frac{\rm e}{2p+1}$ at the point $\zeta_i$,
which means that the quasi-holes are fractionally charged.

Thus, in order to analyze their statistics more carefully, we derive the
\pstress{Berry} connection, first stated
by \pstress{Arovas} et al.~\cite{Arovas:qr},
from the normalization factor (a detailed comment on the derivation is provided
in chapter 2 of \cite{Stone:ft}):
\eqn{&\ds\Psi_{\rm
exc.}={\cal{N}}\ds\prod_{k,l}(z_k-\zeta_l)\prod_{r<s}(z_r-z_s)^{2p+1}
     (\zeta_r-\zeta_s)^{\frac{1}{2p+1}}
      \exp\left[-{\rm F}\left(z_i,\,\zeta_i\right)\right]\:\:
      ,\label{psiexc}\\[1ex]
 &\ds{\rm F}\left(z_i,\,\zeta_i\right)=\frac{1}{4}\ds\sum_i
    \Big(\abs{z_i}^2+\frac{1}{2p+1}
    \abs{\zeta_i}^2\Big)\:\: .\nonumber}
Therefore, the quasi-particles obey fractional statistics and the
non-holomorphic part in the wave function describing quasi-particle
interactions gives rise to the complex geometry the \pstress{Laughlin} states
are built on. This geometrical features are directly embedded in the
\nstress{$b$/$c$-spin} systems. Given a filling fraction $\nu=1/(2p+1)$ we
identify a $\mathbb{Z}_{2p+1}$-symmetric \nstress{projective field} with the
electron ${\rm e}^-$ and another one with the flux quantum $\Phi$,
respectively.

The charge vectors are related to the statistics, thus $(2p+1)$-dimensional and
take the form
\eqn{\vec{\alpha}_{{\rm e}^-}=\Big(\:1,\,\ldots,\,1\:\Big)
     \quad ,\quad
     \vec{\alpha}_{\Phi^{\vphantom{-}}}=
     \Big(\:\frac{1}{2p+1},\,\ldots,\,\frac{1}{2p+1}\:\Big)
     \label{chargevectors}\:\: .}
The correlators \eqref{corr} yield the correct wave functions \eqref{psilaugh}
and \eqref{psiexc} up to the exponential factor:
\eqn{\Psi_{\rm Laughlin}&\!\!\!=\!\!\!&
     \left\langle\,\Omega\left| V_{\vec{\alpha}_{{\rm e}^-}}(z_1)
     \cdot\ldots\cdot
     V_{\vec{\alpha}_{{\rm e}^-}}(z_n)
     \right|0\,\right\rangle\nonumber
     =\prod_{i<l}^n(z_i-z_l)^{2p+1}\:\: ,\\
     \Psi_{\rm exc.}&\!\!\!=\!\!\!&
     \left\langle\,\Omega\left| V_{\vec{\alpha}_{{\rm e}^-}}(z_1)
     \cdot\ldots\cdot
     V_{\vec{\alpha}_{{\rm e}^-}}(z_n)
     V_{\vec{\alpha}_{\Phi^{\vphantom{-}}}}(\zeta_1)\cdot\ldots\cdot
     V_{\vec{\alpha}_{\Phi^{\vphantom{-}}}}(\zeta_k)
     \right|0\,\right\rangle\nonumber\\
     &\!\!\!=\!\!\!&\prod_{r,s}^{n,k}(z_r-\zeta_s)\prod_{i<l}^n(z_i-z_l)^{2p+1}
     \prod_{p<q}^k(\zeta_p-\zeta_q)^{\frac{1}{2p+1}}\:\: .\label{firstjain}}
We have to make a comment here: In our approach, the CFT always lives on a
ramified covering of the compactified complex plane, i.e., on the
\pstress{Riemann} sphere. On the other hand, the FQH system lives on a certain
chunk of the plane, the sample. Thus, in a correct treatment, wave functions of
the FQH system must be elements of a suitable test space. It turns out that
this is the \pstress{Bargmann} space \cite{Girvin:1984aa}.
The elements of the \pstress{Bargmann}
space for $N$ complex variables are of the form
\[\psi(\{z\})=p(z_1,\,\ldots,\, z_N)\prod_{i=1}^N
\exp(-c_i|z_i|^2)\:\:.\] There are further restrictions on the constants $c_i$
and on the multi-variate polynomial $p(\{z\})$ whenever the function
$\psi(\{z\})$ is symmetric or anti-symmetric under certain permutations of its
arguments. The only effect of the exponential factor is to guarantee a
sufficient fast decay of the modulus squared of the wave function if one or
more of its arguments become large. It can be shown rigorously that this factor
is absent if the FQH problem is considered in a different setting, i.e., on a
sphere pierced by the field of a magnetic monopole positioned in its centre.
This idea was first stated by \pstress{Haldane} \cite{Haldane:1983xm}. Since
this is a compact space, so is the support of the wave function. When computing
bulk wave functions in terms of CFT correlators, we automatically move to this
latter setting on the compact sphere. Thus, it is natural to expect that the
CFT picture reproduces the bulk wave functions on the sphere and not on the
plane. However, for completeness, we mention that it is possible to reproduce
the exponential factors within the CFT picture by explicitly including a
homogeneous background charge distribution confining the support of the wave
function as it was shown by \pstress{Moore} and \pstress{Read}
\cite{Moore:ks,Read:1991bt}.

We can deduce that the $\mathbb{Z}_n$-symmetry of the RS the spin fields live
on has a one-to-one correspondence with the statistics and charges of the
(quasi-)particles in the \pstress{Laughlin} states, e.g., the
$(2p+1)$-dimensional charge vectors \eqref{chargevectors} yield the wave
functions \eqref{firstjain}, and $n=2p+1$. Furthermore, the scalar products of
the charge vectors determine the particles' interaction, i.e., order of zeros
in the polynomial terms of the wave functions. We stress that in spite of the
electron with elementary charge ${\rm e}$ obeying simple fermionic statistics
the field's nature has a geometric background in terms of the topology of the
RS. This will become more apparent in states of higher order.
\subsection*{Beyond Laughlin}
As already pointed out the FQHE is a \nstress{strongly correlated system}
(SCS). In such systems interactions dominate the physics and long range effects
take place. Well known examples are superconductivity and the \pstress{Hubbard}
model which can be described in terms of effective theories. The common feature
of these theories is the demand for the existence of effective particles in the
system, e.g., \pstress{Cooper} \nstress{pairs} (superconductivity) or
\nstress{spinons} and \nstress{holons} (\pstress{Hubbard} model). Concerning
the FQHE one widely accepted effective theory with direct correspondence to
experimental facts was developed by \pstress{Jain}
\cite{Jain:1989tx,Jain:1990bb,Jain:1990aa}.
He explained the fractional
effect by introducing the \nstress{composite fermion} (CF) model. A CF consists
of one electron with a number of pairs of flux quanta of the magnetic field
attached to it. \pstress{Jain} showed that the FQHE is an effective IQHE for
the CFs and proposed sequences of states to appear in a
certain order. These are found in agreement with experimental data.

With respect to trial wave functions the attachment of $p$ pairs of flux quanta
is conducted by multiplying the IQHE wave function $\Psi_{\rm I}$ (filling
fraction $\nu_{\rm I}$) with a polynomial \pstress{Jastrow} factor
\eqn{\Psi_{\rm CF}=\prod_{i<j}^{N}(z_i-z_j)^{2p}\Psi_{\rm I}\:\:
.\label{jainwave}} The filling fraction of the CF state is then derived as
\eqn{\nu_{\rm CF}=\frac{\nu_{\rm I}}{2p\nu_{\rm I}+1}\:\: .}
Here, $\nu_{\rm I}$ corresponds to the IQH state $\Psi_{\rm I}$. This procedure
neither destroys the correlations of the system nor the incompressibility of
the state. \pstress{Laughlin}'s wave functions are the simplest examples of
this scheme. We start from a $\nu=1$ IQH state $\Psi_1$  and attach $p$ pairs
of flux quanta:
\eqn{\Psi_{\rm Laughlin}=\prod_{i<j}^N(z_i-z_j)^{2p}
     \underbrace{\prod_{i<j}^N(z_i-z_j)
     \exp\Big(-\frac{1}{4}\sum_i\abs{z_i}^2 \Big)}_{\ds\Psi_1},
     \:\: \nu=\frac{1}{2p+1}\:\: .}
In principle, it is possible to get any rational number as filling factor by
applying \pstress{Jain}'s construction repeatedly. This forms the hierarchical
scheme of \pstress{Jain}. Thus, instead of starting with an IQH state, one
starts with a FQH state obtained from \pstress{Jain}'s construction, and forms
new CFs out of the old ones by attaching additional pairs of flux quanta. The
new filling fraction is obtained via (46) by replacing $\nu_{{\rm I}}$ by
$\nu_{{\rm CF}}$ to obtain a new filling $\nu_{{\rm CF}}'$. In this way,
arbitrarily continued fractions of the form
\def\dfrac#1#2{{\displaystyle\frac{#1}{#2}}}
\begin{equation}
  \nu=[2p_1,2p_2,\,\ldots,\, 2p_n,\nu_{{\rm I}}] =
  \dfrac{1}{2p_1 +
    \dfrac{1}{2p_2 +
      \dfrac{1}{\ddots \raisebox{-8pt}{$\,2p_n+
        \dfrac{1}{\nu_{{\rm I}}}$}}}}
\end{equation}
can be constructed, and thus arbitrary positive rational numbers $\nu<1$.
However, this hierarchical scheme shares with all the other hierarchical
schemes that it soon produces way too many unobserved filling fractions.
Moreover, it is necessary to invoke the principle of particle-hole duality in
order to get some of the experimentally confirmed filling fractions within the
first few levels of the hierarchy. Unfortunately, the set of all experimentally
observed FQH states does not support particle-hole duality very well. Thus, we
avoid this principle in our approach.
%
%  Section - Multilayer States and CFT Approach
%
\section{Multilayer States and CFT Approach}
We demonstrate that particle-hole duality is not needed and, without predicting
unobserved fractions, we derive sequences of all FQH states ($0\leq\nu\leq 1$)
observed up to now in agreement with experimental data (see for example
\cite{Stormer:1999aa,Stormer:2000aa,Stormer:2001aa,Stormer:2002aa}) with a very
few exceptions. Furthermore, a unifying scheme for the construction of bulk
wave functions in terms of CFT correlators is provided. To arrive at these wave
functions we have to generalize \pstress{Jain}'s \nstress{composite fermion}
(CF) approach to multilayer states. One way to provide this is to start from an
effective field theory.

It is well-known that QED in (2$+$1) dimensions consists of a \pstress{Maxwell}
part and a topological \pstress{Chern-Simons} term. It is true that the latter
is neglectable compared to the first one in many cases, but it was rigorously
shown that it dominates the FQH regime \cite{Frohlich:wb}. Therefore, the FQH
system can be described in terms of an effective \pstress{Chern-Simons} theory.
It turns out that a FQH system can consist of several quantum fluids which may
be coupled to each other. Each fluid $i$ is described in the effective field
theory by a vector potential $a_i^{\mu}$ in addition to the external field
$A^{\mu}$ with couplings $\kappa_i$. For completeness, we provide the general
form of the \pstress{Lagrangian}:
\begin{equation}\mathcal{L}=-\frac{1}{4\pi}
     a_{i\mu}K_{ij}\epsilon^{\mu\nu\lambda}\partial_\nu a_{j\lambda}
     -\frac{\rm e}{2\pi}\kappa_iA_\mu\epsilon^{\mu\nu\lambda}\partial_\nu
     a_{i\lambda}
%    +s_i\omega_k\epsilon^{k\nu\lambda}\partial_\nu a_{i\lambda}
%    +l_ia_{i\mu}j^\mu
     +\ldots\:\:,
\end{equation}
where we left out possible other terms such as the contribution of the
quasi-hole current. A very detailed approach is given in \cite{Wen:qn,Wen:uk}.
The complete \pstress{Lagrangian} contains various couplings and sources which
are irrelevant for our purposes. The only important conclusion in our context
is that the internal structure of a so-called $m$-layer FQH state is encoded in
the invertible $m\times m$ matrix $K_{ij}$ describing the couplings of
different layers or quantum fluids with each other. This matrix encodes various
information of the FQH state, e.g.~, the filling fraction, topological order,
ground state degeneracy and the structure of corresponding trial wave
functions. As a result, for an electron system $K_{ij}$ has to satisfy the
following conditions (represented in the symmetric electron basis of
\pstress{Chern-Simons} theory):
\eqn{K_{ij}=\left\{
       \begin{array}{ll}
         \text{odd integer} & i=j\\
     \text{integer} & i\neq j
       \end{array}
       \right.\:\: .}
The filling fraction is
\eqn{\nu_K=\sum_{i,j}^mK^{-1}_{ij}\:\: .}
In addition, the trial wave functions can be read off directly:
\eqn{\Psi_K=\prod_{i<j}^N\prod_\mu^m(z_i^{(\mu)}-z_j^{(\mu)})^{K_{\mu\mu}}
     \prod_{i,j}^N\prod_{\mu<\lambda}^m
     (z_i^{(\mu)}-z_j^{(\lambda)})^{K_{\mu\lambda}}\:\: .\label{kwave}}
\subsection*{Jain's Main Series}
To follow \pstress{Jain}'s approach \eqref{jainwave} we start from a
double-layer IQH state $\Psi_{\rm I}$ with two filled \pstress{Landau} levels
(LLs):
\eqn{K_{ij}=\left(\begin{array}{cc}
     1&0\\0&1
     \end{array}\right)\, ,\quad \nu=2\:\: .}
These two layers do not interact. Attaching $p$ pairs of flux quanta to each
electron yields
\eqn{K_{ij}=\left(\begin{array}{cc}
     2p+1&2p\\2p&2p+1
     \end{array}\right)\, ,\quad \nu=\frac{2}{4p+1}\:\: .}
The flux quanta introduce interactions between different layers. Two filled LLs
of CFs correspond to a LLL FQH state. Generalized to $m$ layers we obtain
\eqn{K_{ij}=\left(\begin{array}{cccccc}
     2p+1&~2p~&~~\cdots~~&\cdots&~2p~\\[1.5ex]
     2p&2p+1&\ddots&&\vdots\\[1.5ex]
     \vdots&\ddots&\ddots&\ddots&\vdots\\[1.5ex]
     \vdots&&\ddots&2p+1&2p\\[1.5ex]
     2p&\cdots&\cdots&~2p~&2p+1\\[1.5ex]
     \end{array}\right)\, ,\quad \nu_p=\frac{m}{2mp+1}\:\: .\label{kbasic}}
This implies the following sequences of filling fractions:
\eqn{\label{jainsimple}
\nu_1&\!\!\!=\!\!\!&\frac{1}{3},~\frac{2}{5}
,~\frac{3}{7},~\frac{4}{9},~\frac{5}{11},~\frac{6}{13},~\frac{7}{15}
,~\frac{8}{17},~\frac{9}{19},~\frac{10}{21},\,\ldots\nonumber\\[0.3ex]
\nu_2&\!\!\!=\!\!\!&\frac{1}{5},~\frac{2}{9},~\frac{3}{13},
      ~\frac{4}{17},~\frac{5}{21},~\frac{6}{25},\,\ldots\nonumber\\[0.3ex]
\nu_3&\!\!\!=\!\!\!&\frac{1}{7},~\frac{2}{13},~\frac{3}{19},\,\ldots\\[0.3ex]
\nu_4&\!\!\!=\!\!\!&\frac{1}{9},~\frac{2}{17},\,\ldots\nonumber\\[-1ex]
&\!\!\!\vdots\!\!\!& \nonumber}
These are limited by the \pstress{Wigner}
crystal regime for $\nu\rightarrow 0$ depending on the quality of the sample.
Therefore, the series for $p\geq 5$ were still not observed. On the other hand
we have a cutoff if $m$, the number of LLs of CFs building the state, is
increased. In terms of an effective IQHE this corresponds to the classical
limit $B_{\rm eff}\rightarrow 0$.

The trial wave functions \eqref{kwave} are LLL projections of the true FQH
states. To do the projection properly CFs of different LLs labelled by $(\mu)$
have to be distinguished. The resulting wave function is anti-symmetric only
within each LL, anti-symmetrization over different LLs is unphysical and would
yield a vanishing $\Psi_{\rm K}$ in most cases.

The complete set of states for the sequences \eqref{jainsimple} is included in
the \nstress{$b$/$c$-spin} system approach, (quasi-)particles, their charges
and statistics are described in terms of $\mathbb{Z}_{2mp+1}$-symmetric
\nstress{projective fields}. As before, $p$ labels the number of pairs of flux
quanta attached to the electron and $m$ is the number of filled CF LLs. Each
layer $\mu\in\{1,\,\ldots,\, m\}$ is connected with a $(2mp+1)$-dimensional
charge vector:
\eqn{\vec{\alpha}^{(\mu)}_i=
     \left\{
       \begin{array}{ll}
         1 & 1\leq i\leq 2p\\
         1 & i=2mp+2-\mu\\
         0 & \text{otherwise}
       \end{array}
     \right.\quad .\label{chargebasic}}
This yields
\eqn{\vec{\alpha}^{(\mu)}\cdot\vec{\alpha}^{(\lambda)}=2p+
     \delta_{\mu,\,\lambda}\:\: .}
Naively one might have expected a $(2p+1)m$-dimensional charge vector for an
$m$-layer state. However, this would mean that the flux quanta were independent
for each layer. Identifying these or, equivalently, the base spaces of the $m$
copies of the ramified complex plane immediately leads to
$(2p+1)m-(m-1)=2mp+1$. The correlators \eqref{corr} are hence derived to read
\eqn{\Psi_{p,\, m}(z_i^{(\mu)})&\!\!\!=\!\!\!&
     \langle\,\Omega\,|\prod_\mu^m V_{\vec{\alpha}^{(\mu)}}(z_1^{(\mu)})
     \cdot\ldots\cdot
     V_{\vec{\alpha}^{(\mu)}}(z_N^{(\mu)})
     |\,0\,\rangle \nonumber\\
     &\!\!\!=\!\!\!&
     \prod_{i<j}^N\prod_\mu^m(z_i^{(\mu)}-z_j^{(\mu)})^{2p+1}
     \prod_{i,j}^N\prod_{\mu<\lambda}^m
     (z_i^{(\mu)}-z_j^{(\lambda)})^{2p}\:\: .\label{jainseries}}
Equation \eqref{jainseries} generalizes the result of \eqref{firstjain} and the
basic \pstress{Jain} series \eqref{jainsimple} with $\nu_p=\frac{m}{2mp+1}$ are
identified.
\subsection*{Composite Fermion Pairing}
Concerning other filling fractions all known hierarchical systems, e.g.
\cite{Halperin:fn,Haldane:1983xm,Jain:1990aa,Blok:1990an,Blok:1990nd}, invoke
the principle of \nstress{particle-hole duality}, relating, for example the
series
\eqn{\nu_1&\!\!\!=\!\!\!&\frac{1}{3},~\frac{2}{5}
,~\frac{3}{7},~\frac{4}{9},~\frac{5}{11},~\frac{6}{13},~\frac{7}{15}
,~\frac{8}{17},~\frac{9}{19},~\frac{10}{21},\,\ldots} and
\eqn{\nu^{(1)}_1&\!\!\!=\!\!\!&\frac{2}{3},~\frac{3}{5}
,~\frac{4}{7},~\frac{5}{9},~\frac{6}{11},~\frac{7}{13},~\frac{8}{15}
,~\frac{9}{17},~\frac{10}{19},\,\ldots\:\: .}
The latter is of type $\nu^{(1)}_p=\frac{m}{2mp-1}$ and can be
represented in terms of $m$-layer $K$-matrices
\eqn{K_{ij}=
     \left\{
       \begin{array}{ll}
         2p-1 & i=j\\
         2p & i\neq j
       \end{array}
     \right.\quad .\label{Kprime}}
This does not suit our approach: it demands the existence of charge vectors
$\vec{\alpha}$ and $\vec{\beta}$ corresponding to different layers with
\eqn{\vec{\alpha}^{\, 2}=\vec{\beta}^{\, 2}=2p-1\quad \mbox{and}\quad
\vec{\alpha}\cdot\vec{\beta}=2p\:\:\lightning\:\: .} This is not possible since
it contradicts \pstress{Schwarz}' inequality and indicates that these `dual'
series possess completely new physical features. The analytic structure of the
wave function \eqref{kwave} for $K$-matrices \eqref{Kprime} exhibits that CFs
living in the same layer repulse each other with the power of $(2p-1)$ while
those of different layers repulse themselves by $2p$. This suggests the
existence of an effectively attractive CF interaction within a LL, i.e.
pairing. This is induced by the ${\rm c}=-2$ logarithmic \nstress{$b$/$c$-spin
system} with spin $j=1$ as it was proven for the famous
\pstress{Haldane-Rezayi} state with filling fraction $\nu=5/2$
\cite{Wen:ha,Milovanovic:1996nj,Moore:ks,Flohr:1997aa}.

In analogy to \eqref{bcbosonized} the fields $b(z)$ and $c(z)$ can be bosonized
on a ramified covering of the compactified complex plane locally representing
the ${\mathbb{Z}}_n$-symmetric RS in terms of \nstress{Vertex operators}:
\eqn{
     \begin{array}{ccc}
     b_{\vec{\gamma}}(z)\equals\no{\exp\big(+\ci\vec{\gamma}\vec{\Phi}(z)\big)}
     \\[1ex]
     c_{\vec{\gamma}}(z)\equals\no{\exp\big(-\ci\vec{\gamma}\vec{\Phi}(z)\big)}
     \end{array}&&\gamma_k\in\{0\, ,\, 1\}\:\: .\label{bcpair}
     }
In terms of CFT the pairing effect of the CFs is described by $b(z)\del c(z')$.
The OPE
\eqn{b_{\vec{\gamma}}(z)\del c_{\vec{\gamma}}(z')\sim
     \frac{\vec{\gamma}^{\, 2}}{(z-z')^2}\label{Pfaffbc}}
yields the so-called \pstress{Pfaffian} form ${\rm Pf}(z_i,\, z'_i)$  if the
fields \eqref{Pfaffbc} are evaluated in a correlator:
\eqn{\langle\,\Omega\,|\left(b_{\vec{\gamma}}(z_1)
     \del_{z'_1}c_{\vec{\gamma}}(z'_1)\right)
     \cdot\ldots\cdot
     \left(
     b_{\vec{\gamma}}(z_{N})
     \del_{z'_N}c_{\vec{\gamma}}(z'_N)
     \right)
     |\,0\,\rangle\equals\vec{\gamma}^{\, 2}{\rm Pf}(z_i,\, z'_i)\:\:
     ,\nonumber}
\eqn{{\rm Pf}(z_i,\, z'_i)&\!\!\!\equiv\!\!\!&
     \sum_{\sigma\in S_N}\prod_{i=1}^N
     \frac{1}{(z^{\phantom{\prime}}_i-z^{\prime}_{\sigma(i)})^2}\:\: .}
In this way, the $\nu^{(1)}_p$ series can be identified by the same fields as
the basic \pstress{Jain} series \eqref{jainseries} if additional inner-LL
pairings are included. To find a physical and stable system we expect all CF
LLs to be paired. In order to describe this in a proper way, each layer
$\mu\in\{1, \ldots\ , m\}$ possesses an $m$-dimensional charge vector:
\eqn{\vec{\gamma}^{(\mu)}_i=\delta_{\mu,\, i}\:\:\df\:\:
     \vec{\gamma}^{(\mu)}\cdot\vec{\gamma}^{(\lambda)}=
     \delta_{\mu,\,\lambda}\:\: .}
The CFs themselves correspond to the charge vectors \eqref{chargebasic}. Thus,
the wave function reads
\eqn{\Psi^{(1)}_{p,\, m}(z_i^{(\mu)})&\!\!\!=\!\!\!&
     \langle\,\Omega\,|\prod_\mu^m V_{\vec{\alpha}^{(\mu)}}(z_1^{(\mu)})
     \cdot\ldots\cdot
     V_{\vec{\alpha}^{(\mu)}}(z_{2N}^{(\mu)})
     |\,0\,\rangle\times \nonumber\\
     \hspace{-2cm}&\hspace{-3.7cm}\!\!\!\times\!\!\!&\hspace{-2cm}
     \langle\,\Omega\,|\prod_\mu^m
     \left(b_{\vec{\gamma}^{(\mu)}}(z_1^{(\mu)})
     \del_{z_{N+1}}c_{\vec{\gamma}^{(\mu)}}(z_{N+1}^{(\mu)})\right)
     \cdot\ldots\cdot
     \left(
     b_{\vec{\gamma}^{(\mu)}}(z_{N}^{(\mu)})
     \del_{z_{2N}}c_{\vec{\gamma}^{(\mu)}}(z_{2N}^{(\mu)})
     \right)
     |\,0\,\rangle \nonumber\\
     \hspace{-2cm}&\hspace{-3.7cm}\!\!\!=\!\!\!&\hspace{-2cm}
     \prod_\mu^m{\rm Pf}(z_i^{(\mu)},\, z_{N+i}^{(\mu)})
     \underbrace{\prod_{i<j}^{2N}
     \prod_\mu^m(z_i^{(\mu)}-z_j^{(\mu)})^{2p+1}}_{\left(\star\right)}
     \prod_{i,j}^{2N}\prod_{\mu<\lambda}^m
     (z_i^{(\mu)}-z_j^{(\lambda)})^{2p}
     \:\: .\label{firstpaired}}
We want to stress that equation \eqref{firstpaired} satisfies the
\pstress{Chern-Simons} approach and has to be identified with the $K$-matrix
\eqref{Kprime}. Only the trial wave functions \eqref{kwave} have to be
extended, since they are not capable to realize pairing effects in a proper
way. However, the \pstress{Pfaffian} cancels two powers of the paired CF
contribution to $(\star)$. Thus, paired CFs repulse each other by
$(z_i^{(\mu)}-z_j^{(\mu)})^{2p-1}$ in either wave function. Additionally, both
yield the same filling fractions
\eqn{\nu^{(1)}_{p,\, m}=\frac{m}{2mp-1}\:\: .}
We identify the first order paired series:
\eqn{ \label{jainfirst}
\nu^{(1)}_1&\!\!\!=\!\!\!&\frac{2}{3}
     ,~\frac{3}{5},~\frac{4}{7},~\frac{5}{9},~
     \frac{6}{11},~\frac{7}{13},~\frac{8}{15},~
     \frac{9}{17},~\frac{10}{19},\,\ldots\nonumber\\[0.3ex]
\nu^{(1)}_2&\!\!\!=\!\!\!&\frac{1}{3},~
     \frac{2}{7},~\frac{3}{11},~\frac{4}{15},~\frac{5}{19},
     ~\frac{6}{23},\,\ldots\nonumber\\[0.3ex]
\nu^{(1)}_3&\!\!\!=\!\!\!&\frac{1}{5},~\frac{2}{11},
            ~\frac{3}{17},\,\ldots\\[0.3ex]
\nu^{(1)}_4&\!\!\!=\!\!\!&\frac{1}{7},~\frac{2}{15},\,\ldots\nonumber\\[-1ex]
&\vdots\nonumber} This proposal can be extended in a natural way imagining that
the structure of paired CF singlets is not restricted to be an inner-LL effect.
Two LLs of CFs that are completely paired among each other can form a new
incompressible quantum liquid and can hence interact with other blocks or
single layers of paired droplets. Therefore, we find two natural series of
$K$-matrices $(K^{\rm e,\, o})_{ij}$ with an even and an odd number of layers,
respectively:
\eqn{(K^{\rm e,\, o})_{ij}\equals
     \left\{
       \begin{array}{clcll@{\leq}l@{\,\leq\,}l}
         2p-1 & i\equals j\\
         2p-2 & i&\!\!\!\neq\!\!\!&j,&2(k-1)+1&\, i,j&2k
     \quad (1\leq k\leq b)\\
     2p   & \multicolumn{4}{l}{\mbox{otherwise}}
       \end{array}
     \right.\:\: .}
Here, $b$ is the number of paired $2\times 2$-blocks. The first series, given a
$2b$-layer FQH state, reads:
\eqn{(K^{\rm e})_{ij}\equals\left(\begin{array}{cccccc}
     2p-1&2p-2&~2p~&\cdots&~2p~\\[1.5ex]
     2p-2&2p-1&~2p~&\ddots&\vdots\\[1.5ex]
     ~2p~&~2p~&\ddots&~2p~&~2p~\\[1.5ex]
     \vdots&\ddots&~2p~&2p-1&2p-2\\[1.5ex]
     2p&\cdots&~2p~&2p-2&2p-1\\[1.5ex]
     \end{array}\right)\, ,\:\: \nu^{(2)\,\rm e}_p=\frac{2b}{4bp-3}\:\: .}
The latter, given a $2b+1$-layer FQH state, has a remaining solely self-paired
layer and corresponds to filling fractions
\eqn{\nu^{(2)\,\rm o}_p=\frac{2b+3}{2p(2b+3)-3}\:\: .}
Together, they yield the second order paired series\footnote{Fractions in
brackets are not coprime and also appear in other series. This indicates that
these states can exist in different forms of quantum liquids.}:
\eqn{
\begin{array}{l@{\: =\:}l@{\hspace*{1.5cm}}l@{\: =\:}l}
\nu^{(2)\,\rm e}_1&\displaystyle\frac{4}{5}
     ,\left(\frac{6}{9}\right),~\frac{8}{13},~\frac{10}{17},\,\ldots\:\: ,&
\nu^{(2)\,\rm o}_1&\displaystyle\frac{5}{7},~\frac{7}{11},
     \left(\frac{9}{15}\right),\,\ldots\:\: ,\\[2ex]
\nu^{(2)\,\rm e}_2&\displaystyle\frac{2}{5},~\frac{4}{13},\,\ldots\:\: ,&
\nu^{(2)\,\rm o}_2&\ldots\:\: .
\end{array}\label{jainsecond}
} We are now able to generalize this scheme to the case of $n\times n$ blocks
of paired LLs and derive the $n$-th order series. There exist $n-1$ sub-series
determined by the number $r$ of remaining solely self-paired LLs, e.g.~$r=0$ in
the even case for second order and $r=1$ in the odd case, respectively. Let $b$
denote the number of fully paired blocks then the $m\times m$-matrix
$K^{(n)}_{p,\, m}$ with $m=bn+r$ of the $n$-th order paired FQH state reads:
\eqn{\left(K^{(n)}_{p,\, m}\right)_{ij}\equals\left\{
       \begin{array}{clcll@{\leq}l@{\,\leq\,}l}
         2p-1 & i\equals j\\
         2p-2 & i&\!\!\!\neq\!\!\!&j,&(k-1)n+1&\, i,j&kn
     \quad (1\leq k\leq b)\\
     2p   & \multicolumn{4}{l}{\mbox{otherwise}}
       \end{array}
     \right.\:\: .\label{Kordern}}
The corresponding filling fractions are
\eqn{
\nu^{(n)\, r}_{\phantom{(}p,\, m}
   =\frac{bn+r(2n-1)}{2p(bn+r(2n-1))-(2n-1)}\:\: .\label{nugeneral}}
By this, we deduce the third order states confirmed by experiment\footnote{
  The state
  $\nu=\frac{6}{7}$ has not been confirmed so far, since it falls in the
  domain of attraction of the $\nu=1$ plateau, but is strongly expected.} (higher
  orders do not yield additional observed fractions):
\eqn{
  \begin{array}{l@{\: =\:}l@{\hspace*{1.0cm}}l@{\: =\:}l@{\hspace*{1.0cm}}l@{\:
                =\:}l}
  \nu^{(3)\, 0}_1&\displaystyle\left[\frac{6}{7}\right]
       ,\frac{9}{13},\,\ldots&
  \nu^{(3)\, 1}_1&\displaystyle\frac{8}{11},\,\ldots&
  \nu^{(3)\, 2}_1&\displaystyle\ldots\\[2ex]
  \nu^{(3)\, 0}_2&\displaystyle\frac{3}{7}
       ,\,\ldots&
  \nu^{(3)\, 1}_2&\displaystyle\ldots&
  \nu^{(3)\, 2}_2&\displaystyle\ldots
  \end{array}\label{jainthird}
} Spending a closer look on \eqref{Kordern} the question arises to what extent
our access to FQH pairing is too restrictive. One could imagine more general
$K$-matrices with band-like or even more complicated structures yielding
arbitrary $\nu$. For example, $\nu=4/11$, a state that was very recently
confirmed by experiment \cite{Stormer:2002aa}, could be realized by
\eqn{K_{ij}=\left(\begin{array}{cccc}
     3&2&2&4\\2&3&4&2\\2&4&3&2\\4&2&2&3
     \end{array}\right)\:\: .}
This $K$-matrix describes a ring of two second order blocks.
Remarkably, the result of a detailed analysis of equation \eqref{nugeneral}
shows that certain fractions do not appear,
for example $7/9$, $10/13$, $5/13$, and
$4/11$. In agreement, as far as we know, there merely exist
controversial data concerning the first three,
indicating that if they exist they
presumably have to be another kind of FQH fluid. The same holds for $\nu=4/11$
that is assumed to be a non-\pstress{Abelian} state. As exactly these fractions
lie beyond the access of our scheme,
the \nstress{$b$/$c$-spin} systems motivate a reasonable
physical constraint for the \pstress{Chern-Simons} formalism in order to
classify FQH states. We directly
deduce this from the CFT picture of the fields given by \eqref{bcpair}. If we
had an off-block pairing structure, there would exist a triple of
\eqn{b_{\vec{\gamma_1}}(z_i^{(1)})\del c_{\vec{\gamma_1}}(z_j^{(1)})\:\: ,\:\:
     b_{\vec{\gamma_2}}(z_i^{(2)})\del c_{\vec{\gamma_2}}(z_j^{(2)})\:\: ,\:\:
     b_{\vec{\gamma_3}}(z_i^{(3)})\del c_{\vec{\gamma_3}}(z_j^{(3)})\:\: ,
     }
with the charge vectors obeying the following set of equations:
\eqn{\vec{\gamma_1}^{\, 2}=\vec{\gamma_2}^{\, 2}=\vec{\gamma_3}^{\, 2}=1\quad ,
     \quad\vec{\gamma_1}\cdot\vec{\gamma_2}=\vec{\gamma_1}\cdot\vec{\gamma_3}=1
     \quad\mbox{and}\quad\vec{\gamma_2}\cdot\vec{\gamma_3}=0\:\: .}
Since their components are restricted to be either $0$ or $1$,
we end up with a contradiction:
\eqn{\vec{\gamma_1}=\vec{\gamma_2}=\vec{\gamma_3}\quad\mbox{and}\quad
     \vec{\gamma_2}\neq\vec{\gamma_3}\quad{\lightning}\:\: .}
As a consequence the most general $K$-matrix for a correct description of
paired FQH states is restricted to be built from blocks:
\eqn{\left(K^{b,\,n_b}_{p,\, m}\right)_{ij}\equals\left\{
       \begin{array}{clcll@{\leq}l@{\,\leq\,}l}
         2p-1 & i\equals j\\[-1.5ex]
         2p-2 & i&\!\!\!\neq\!\!\!&j,&1\!+\!\!\ds\sum_{l=1}^{k-1}n_l\,&\, i,j&
     \ds\sum_{l=1}^{k}n_l
         \quad (1\leq k\leq b)\\[-1.5ex]
         2p   & \multicolumn{4}{l}{\mbox{otherwise}}
       \end{array}
     \right.\:\: .\label{Kmaxgen}}
Here, $b$ denotes the number of blocks and $n_b$ their corresponding size.
Therefore, $m=\sum_{l=1}^{b}n_b\,$, if we denote singly paired layer by
$n_b=1$. We stress that the new series of filling fractions $\nu^{b,\, n_b}_p$
obtained from \eqref{Kmaxgen} are rather unlikely to be seen in experiments as
their $K$-matrices are less symmetric than the ones given by \eqref{Kordern}.
Since it is quite difficult to derive a general formula for
$\nu^{b,\, n_b}_p\,$, we simply provide the only additional fraction that may
be seen in the nearer future:
\eqn{\nu^{2,\,(3,2)}_1=\frac{19}{23}\:\: .}
Therefore, the set of matrices \eqref{Kordern} remains as the natural candidate
to describe series of paired FQH states by order of stability. The
corresponding bulk wave functions $\Psi^{(n)}_{p,\, m}$ of the $n$-th order
paired FQH states can be calculated as a direct generalization of
\eqref{firstpaired}. Given the matrix $K^{(n)}_{p,\, m}\,$, an $m$-dimensional
charge vector with respect to a paired block $B\in\{1,\,\ldots,\, b+r\}$
(either $n\times n$ or a remaining $1\times 1$ layer) is identified with each
layer $\mu$ :
\eqn{\vec{\gamma}^{(\mu)}_i=\delta_{B(\mu),\, i}\:\:\df\:\:
     \vec{\gamma}^{(\mu)}\cdot\vec{\gamma}^{(\lambda)}=
     \delta_{B(\mu),\, B(\lambda)}\:\:.}
Additionally, each layer $k$ possesses
a $(2mp+1)$-dimensional charge vector for the CFs:
\eqn{\vec{\alpha}^{(\mu)}_i=
     \left\{
       \begin{array}{ll}
         1 & 1\leq i\leq 2p\\
         1 & i=2mp+2-\mu\\
         0 & \text{otherwise}
       \end{array}
     \right.\quad \df\quad
     \vec{\alpha}^{(\mu)}\cdot\vec{\alpha}^{(\lambda)}=
     2p+\delta_{\mu,\,\lambda}\:\: .
     \label{zncharge}}
Let $I$ denote the set of paired LLs, e.g., $I=\{(1,1),\, (2,2),\, (3,3),\,
(1,2),\, (2,1)\}$ describes a triple-layer state with $\nu^{(2)\,
1}_{\phantom{(}p,\, 3}=\textstyle\frac{5}{10p-3}$ where we find a $2\times
2$-block of the first two LLs while the third is solely self-paired. 
The wave functions read
\eqn{\Psi^{(n)}_{p,\, m}(z_i^{(\mu)})&\!\!\!=\!\!\!&
     \langle\,\Omega\,|\prod_\mu^m V_{\vec{\alpha}^{(\mu)}}(z_1^{(\mu)})
     \cdot\ldots\cdot
     V_{\vec{\alpha}^{(\mu)}}(z_{2N}^{(\mu)})
     |\,0\,\rangle\times \nonumber\\
     \hspace{-2cm}&\hspace{-3.7cm}\!\!\!\times\!\!\!&\hspace{-2cm}
     \langle\,\Omega\,|\!\!\!\!\prod_{(\mu,\,\lambda)\in I}\!\!\!\!
     \big(b_{\vec{\gamma}^{(\mu)}}(z_1^{(\mu)})
     \del_{z_{N+1}}c_{\vec{\gamma}^{(\lambda)}}(z_{N+1}^{(\lambda)})\big)
     \cdot\ldots\cdot
     \big(
     b_{\vec{\gamma}^{(\mu)}}(z_{N}^{(\mu)})
     \del_{z_{2N}}c_{\vec{\gamma}^{(\lambda)}}(z_{2N}^{(\lambda)})
     \big)
     |\,0\,\rangle \nonumber\\
     \hspace{-2cm}&\hspace{-3.7cm}\!\!\!=\!\!\!&\hspace{-2cm}
     \!\!\prod_{(\mu,\,\lambda)\in I}\!\!\!\!
     {\rm Pf}(z_i^{(\mu)},\, z_{N+i}^{(\lambda)})
     \underbrace{\prod_{i<j}^{2N}
     \prod_\mu^m(z_i^{(\mu)}-z_j^{(\mu)})^{2p+1}
     \prod_{i,j}^{2N}\prod_{\mu<\lambda}^m
     (z_i^{(\mu)}-z_j^{(\lambda)})^{2p}}_{{\textstyle\Psi_{p,\, m}}(
     z_i^{(\mu)})}
     \:\: ,\label{finalwave}}
where $\Psi_{p,\, m}(z_i^{(\mu)})$ is the bulk wave function of the basic
\pstress{Jain} series \eqref{jainseries}.

Combining equations \eqref{jainsimple}, \eqref{jainfirst}, \eqref{jainsecond},
and \eqref{jainthird} we find the complete set\footnote{Except $\nu=4/11$,
which is presumably a non-\pstress{Abelian} FQH state falling
outside our approach, and
controversial fractions as $\nu=7/9$, $\nu=10/13$, and $\nu=5/13$.} of
experimentally confirmed filling fractions by order of stability.
\begin{center}
\hspace*{-1cm}
\epsfig{file=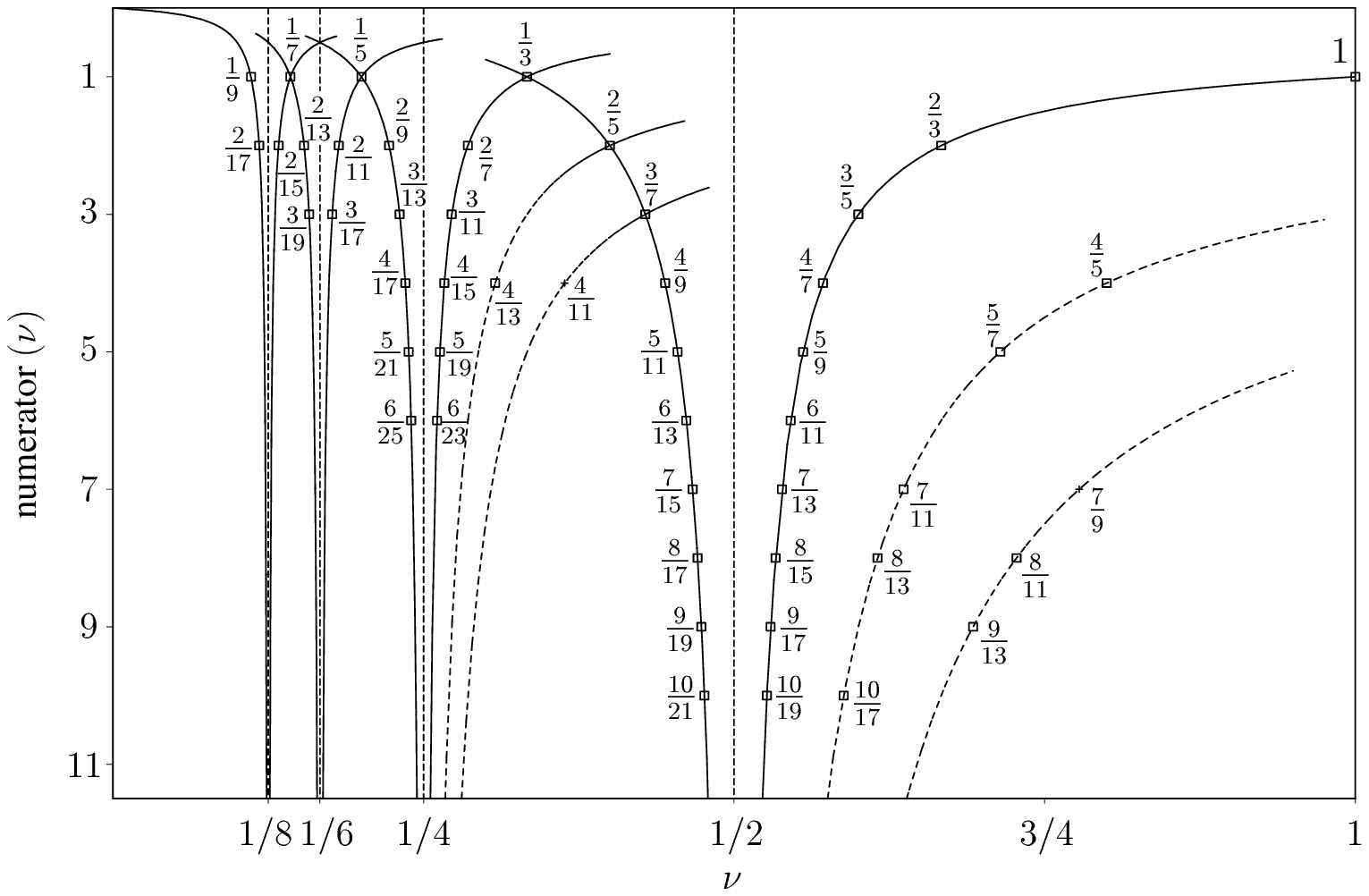,width=16cm} \refstepcounter{figure}
\label{fig:fractions} {Figure
\ref{fig:fractions}:\em\ Observed \pstress{Hall} fractions in the interval
$0\leq\nu\leq 1$}
\end{center}
\rm Established fractions are labelled by `$\Box$'. The symbol
`$\scriptstyle+$' denotes cases that exceed our scheme. The basic
\pstress{Jain} series $\nu_p$ approximate $1/2p$ from below, the corresponding
first order paired series $\nu^{\scriptscriptstyle(1)}_{\scriptscriptstyle p}$
from above (both marked by continuous lines) as well as the higher order series
$\nu^{\scriptscriptstyle(n)}_{\scriptscriptstyle p}$ (marked by dashed lines
).\\[2ex] We find a natural cutoff if either the number of participating CF LLs
$m$ increases or $\nu\rightarrow 0$. Series of more complicated CFs (larger
$p$) are less developed, complete pairings ($r=0$) are favored and each series
precisely keeps track of the stability of the FQH states found in experiments
whereas no unobserved fraction is predicted.

We want to make another comment on the absence of the $\nu=7/9$ state.
If we naively assumed the series
\eqn{\nu=\frac{k}{2k-5}=
\frac{6}{7},~\frac{7}{9},~\frac{8}{11},~\frac{9}{13},\,\ldots\:\: , \nonumber}
we would consider $\nu=7/9$ to be more likely to appear than $\nu=8/11$.
Furthermore, it cannot be argued that $7/9$ is dominated by the $\nu=1$ plateau
since $\nu=4/5$ exists. This seems rather unusual or even exceptional but is
precisely predicted by our approach. Therefore, the series in figure
\ref{fig:fractions} simply indicate where new fractions given by
\eqref{nugeneral} will show up. By our hierarchical order of stability the
following filling fractions are predicted if experimental circumstances are
improved in the future (we just indicate fractions with denominator $d\leq
29$).
\eqn{
\begin{array}{c|c|c|c|c|c|}
     \ds p&\ds\phan\ \nu_p\phan&\ds\phan\nu^{(1)}_p\phan&
     \ds\phan\nu^{(2)}_p\phan&\ds\phan\nu^{(3)}_p\phan&
     \ds\phan\nu^{(4)}_p\phan\\[1ex]\hline
     \ds 1&\ds\frac{11}{23}^{\vphantom{B^B}}&\ds\frac{11}{21}
     &\ds \frac{11}{19},\,\frac{12}{21}&
     \ds \frac{11}{17},\,\frac{12}{19},\,\frac{13}{21}&\ds
     \frac{8}{9},\, \frac{11}{15},\, \frac{18}{29}\\[1.5ex]\hline
     \ds 2&\ds \frac{7}{29}^{\vphantom{B^B}}&\ds \frac{7}{27}&
     \ds \frac{5}{17},\,\frac{6}{21}&
     \ds \frac{6}{19},\, \frac{8}{27}&\ds \frac{8}{25}\\[1.5ex]\hline
     \ds 3&\ds \frac{4}{25}^{\vphantom{B^B}}&\ds \frac{4}{23}&
     \ds \frac{4}{21},\, \frac{5}{27}&&\\[1.5ex]\hline
     \ds 4&\ds \frac{3}{25}^{\vphantom{B^B}}&\ds \frac{3}{23}&
     \ds \frac{4}{29}&&
     \\[1.5ex]\hline
\end{array}\nonumber
}
\begin{center}
{Table 1:\em\ Expected \pstress{Hall} fractions}
\end{center}
\subsection*{Quasi-Particle Excitations}
One of the most striking results in the study of the FQHE was the discovery of
quasi-particles with fractional charges and statistics \cite{Laughlin:1983fy}.
Experimentally it has been proven very difficult to measure them (even for the
\pstress{Laughlin} states) and a lot of effort is spent to analyze them in more
detail. The two sets of wave functions \eqref{jainseries} and \eqref{finalwave}
describe the electron ground state for a given filling fraction $\nu$. As
already shown for the \pstress{Laughlin} series the geometric features of
excitations responsible for statistics and charges are directly embedded in the
\nstress{$b$/$c$-spin} systems and are related to the $\mathbb{Z}_n$-symmetry
of the RS the fields live on, i.e., the dimension of the CF charge vectors
\eqref{zncharge}. However, a basic quasi-particle excitation of an $m$-layer
state has to be considered more carefully. First of all, we would like to have
trivial statistics of a quasi-particle with respect to the CFs. Thus, we should
expect $\vec{\alpha}^{}_{\Phi}\cdot\vec{\alpha}^{(\mu)}_{{\rm CF}} = 1$, where
$\vec{\alpha}^{(\mu)}_{{\rm CF}}$ is the charge vector of the CF in the
$\mu$-th layer as given by \eqref{chargebasic}. The naive solution
\eqn{\vec{\alpha}_{\Phi}=
\Big(\:\frac{1}{2mp+1},\,\ldots,\,\frac{1}{2mp+1}\:\Big)\:\: ,\quad
\left(\vec{\alpha}_{\Phi}\right)^2=\frac{1}{2mp+1}}
yields the value
$\vec{\alpha}^{}_{\Phi}\cdot\vec{\alpha}^{(\mu)}_{{\rm CF}}
= \frac{2p+1}{2pm+1}$, which is not an integer for $m>1$.
A simple generalized solution exists, namely
\begin{equation}
  \vec{\alpha}^{}_{\Phi,i}=\frac{1}{2pm+1}
     \left\{
       \begin{array}{cl}
         m & 1\leq i\leq 2p\\
         1 & 2pm+2-m\leq i\leq 2pm+1\\
         0 & \text{otherwise}
       \end{array}
     \right.\:\: ,\label{chargeqp}
\end{equation}
which coincides with \eqref{chargevectors} for $m=1$.
By this, we obtain the desired result for all layers. Furthermore,
\eqn{\vec{\alpha}^{}_{\Phi}\cdot\vec{\alpha}^{}_{\Phi}=
\frac{1}{(2pm+1)^2}(2pm^2+m) = \frac{m}{2pm+1}\:\: ,} which yields the correct
quasi-particle statistics for an $m$-layer state since each layer contributes
$1/(2pm+1)$.

Thus, the quasi-particle excitations of the wave functions $\Psi_{p,\, m}$ and
$\Psi^{(n)}_{p,\, m}$ are predicted to carry a phase $\Theta\sim\pi/(2mp+1)$
and have the charge $q\sim{\rm e}/(2mp+1)$.
\eqn{
\begin{array}{|c|c||c|c|}\hline
\ds\vphantom{\frac{1}{3}^B}\,\Theta\,&
\ds\vphantom{\frac{1}{3}^B}\hspace*{2.2cm}\nu\hspace*{2.2cm}
             &\,\Theta\,&\ds\hspace*{2.2cm}\nu\hspace*{2.2cm}\\[1.4ex]
          \hline
\ds\frac{\pi}{3}&\ds\vphantom{\frac{1}{3}^B}
                 \frac{1}{3}&
\ds\frac{\pi}{15}&\ds\vphantom{\frac{1}{15}^{B^B}}\frac{7}{15},\,\frac{7}{13},
                  \!\bigg(\frac{9}{15}\bigg)\hspace{-0.2cm}\\[1.6ex]\hline
\ds\frac{\pi}{5}&\ds\vphantom{\frac{1}{5}^B}\frac{1}{5},\,
                 \frac{2}{5},\,\frac{2}{3}&
\ds\frac{\pi}{17}&\ds\vphantom{\frac{1}{17}^B}
                  \frac{2}{17},\,\frac{2}{15},\,
          \frac{4}{17},\,
                  \frac{4}{15},\,\frac{4}{13},\,
                  \frac{4}{9},\,\frac{8}{17},\,\frac{8}{15},\,
          \frac{8}{13}\\[1.4ex]\hline
\ds\frac{\pi}{7}&\ds\vphantom{\frac{1}{7}^B}\frac{1}{7},\,
                 \frac{3}{7},\,\frac{3}{5},\,\frac{5}{7}&
\ds\frac{\pi}{19}&\ds\vphantom{\frac{1}{19}^B}
                  \frac{3}{19},\,\frac{3}{17},\,
          \frac{3}{13},\,\frac{9}{19},\,
                  \frac{9}{17},\,\frac{9}{13}\\[1.4ex]\hline
\ds\frac{\pi}{9}&\ds\vphantom{\frac{1}{9}^B}\frac{1}{9},\, \frac{2}{9},\,
                 \frac{2}{7},\,\frac{2}{5},\,\frac{4}{9},\,
                 \frac{4}{7},\,\frac{4}{5},\,\frac{8}{11}&
\ds\frac{\pi}{21}&\ds\vphantom{\frac{1}{21}^B}
                  \frac{5}{21},\,\frac{5}{19},\,\frac{5}{11},\,
                  \frac{10}{21},\,\frac{10}{19},\,\frac{10}{17}\\[1.4ex]\hline
\ds\frac{\pi}{11}&\ds\vphantom{\frac{1}{11}^B}\frac{5}{11},\,
                  \frac{5}{9},\,\frac{7}{11}&
\ds\frac{\pi}{23}&\ds\vphantom{\frac{1}{23}^B}\\[1.4ex]\hline
\ds\frac{\pi}{13}&\ds\vphantom{\frac{1}{13}^{B^B}}
                  \frac{2}{13},\,\frac{2}{11},\,\frac{3}{13},\,
                  \frac{3}{11},\,\frac{3}{7},\,\frac{6}{13},\,\frac{6}{11},
          \!\bigg(\frac{6}{9}\bigg)\!\!&
\ds\frac{\pi}{25}&\ds\vphantom{\frac{1}{25}^B}
                  \ds \frac{6}{25},\,\frac{6}{23}\\[1.6ex]\hline
\end{array}\nonumber
}
\begin{center}
{Table 2:\em\ Quasi-particle statistics for confirmed FQH states}
\end{center}
Since several filling fractions, e.g.~$2/5$, belong to more than one series
and, thus, exist in different forms of quantum liquids, we also find various
types of quasi-particles. Direct experimental observations are still difficult,
and
--- as far as we know --- good indications solely exist for the
\pstress{Laughlin} series. Thus, the correct identification of the
quasi-particle within the spectrum of our CFT must remain open. We finally note
that our choice (\ref{chargebasic}) for the charge vectors of the CF and
\eqref{chargeqp} for the quasi-particles is not unique, although physically
motivated, particularly simple and symmetric. The ambiguity is not disturbing
since most other solutions are related to ours by a change of basis within the
tensor product of the CFTs. The advantage of our approach is that the CFTs
themselves are confined to a discrete series leaving not much room for
arbitrariness.

%
%  Section - Summary and Outlook
%
\section{Summary and Outlook}
The success of the analysis of the \pstress{Haldane-Rezayi} state via
${\rm c}=-2$ spin systems \cite{Flohr:1997aa,Cappelli:1998ma}
stimulated our approach. With a few general and physically motivated
assumptions restricting to a discrete set of CFTs we were able to construct a
hierarchical scheme that precisely keeps track of experimental results.
Developing these features in a natural and simple way,
we consecutively derived, with a few exceptions,
the complete set of filling fractions by order of stability
in the FQH regime of $0\leq\nu\leq 1$ without predicting fractions
not confirmed by experiment.

More precisely, we constructed CFTs yielding geometrical descriptions of FQH
states. Since odd-denominator fillings refer to fermionic statistics, the
natural choice are $(j,1-j)$ \nstress{$b$/$c$-spin} systems with $j$
half-integer. Moreover, the statistics of the flux quanta, as suggested by
\pstress{Jain}'s composite fermion picture, are now more general such that we
are led to consider \pstress{Riemann} surfaces with global
$\mathbb{Z}_n$-symmetry. Representing these surfaces as $n$-fold ramified
covering of the complex plane, the effect of a flux quantum is geometrically
the same as a branch point. The CFT correlators are then sections of certain
vector bundles. The bulk ground state wave function is given by a correlator of
vertex operators whose twist numbers are purely fermionic resembling the
quantum numbers of a composite fermion. With these ingredients we obtained bulk
wave functions for the principal main series $\nu=\frac{m}{2pm+1}$. It turned
out that our choice of CFTs has not only a direct geometric interpretation, but
furthermore puts severe constraints on possible FQH states. The description of
the FQHE via an effective \pstress{Chern-Simons} theory leads to a
classification of FQH states in terms of the so-called $K$-matrices. Our
approach rules out many $K$-matrices, since the corresponding bulk wave
functions can not be written in factorized form in terms of CFT correlators.

Besides the main series of \pstress{Jain}, we obtain other filling fractions by
one further principle. We point out that within our work we do not use the
so-called particle-hole duality, since it is not well confirmed by experiment.
Instead, we introduce pairings of composite fermions. This leads to a new
hierarchy of states obtained from the principal series by a growing number of
pairings that are effectively described by additional CFTs, namely the already
mentioned ${\rm c}=-2$ spin singlet systems. The requirement that the bulk wave
function can be written in terms of factorized CFT correlators demands that
only pairings leading to $K$-matrices in block form are possible. By this, we
obtain all experimentally observed filling fractions\footnote{Except for
$\nu=4/11$, which is presumably a
non-\pstress{Abelian} FQH state falling outside our
approach, and for controversial fractions as $\nu=7/9$, $\nu=10/13$, and
$\nu=5/13$.}. However, the great strength and predictive power we see in our
approach is that it does precisely avoid all the filling fractions which are
not observed in nature. Our ansatz yields a natural order of stability in
perfect agreement with experimental data suggesting a clear picture of series
which can be observed up to a given maximal numerator of $\nu$. Thus, we are
able to denote the next members of these series, as indicated by figure 1 and
table 1, which might be observed under improved experimental conditions, but no
other fractions.

The main advantage of our scheme is that it avoids arbitrariness and the
concept of pairing is not exceptional as well. First of all, it precisely
agrees with experimental observations for the \pstress{Haldane-Rezayi} state. A
nice discussion is provided by \cite{Read:2001aa}. Moreover, pairing effects
are indicated by numerical studies \cite{Morf:1986aa,Morf:1986bb}, and are in
analogy to similar phenomena in other fields of condensed matter physics, such
as certain exactly integrable models in the context of BCS pairing
\cite{Sierra:1999mp}. Although our proposed bulk wave functions which describe
paired FQH states differ from the ones predicted by the naive $K$-matrix
formalism, they share important asymptotic features. A check of our bulk wave
functions should be done numerically, but is beyond the scope of this paper.

Our description in terms of \nstress{$b$/$c$-spin} systems seems to be
sufficiently complete. It should be possible to incorporate FQH states from
non-\pstress{Abelian Chern-Simons} theories
\cite{Ardonne:2000qg,Ardonne:2001jq} as well, since we believe that the
geometric principle remains unchanged. The main difference lies in the nature
of the quasi-particle excitations. In our approach, non-trivial statistics is a
consequence of the twists introduced by the flux quanta and is -- in the LLL --
always of \pstress{Abelian} nature since all monodromies are simultaneously
diagonalized. Non-\pstress{Abelian} statistics is involved and cannot be
represented within the simple CFTs we used. However, we point out that the
${\rm c}=-2$ CFT coming into play with pairing is actually a logarithmic CFT
and thus includes fields with non-diagonalizable monodromy action
\cite{Flohr:1997aa}. In order to understand this in more detail, we would have
to work with the full twist fields, not only the projective ones. This
immediately leads to further restrictions for the twist fields in order to be
inserted in a correlator. If the twists are summed over all insertions they
have to be trivial in all $n$ copies of the \nstress{$b$/$c$-spin} system we
consider. However, at this stage, the full description of quasi-particle
excitations remains an unsolved problem. Another one is the correct choice of
the spin system, i.e., of the conformal weights $(j,1-j)$ of the field $b(z)$
and $c(z)$. This problem is related to the fact that our \nstress{$b$/$c$-spin}
systems possess partition functions which are equivalent to \pstress{Gaussian}
${\rm c}=1$ models. Unfortunately, the partition function of a $(j,1-j)$ system
is closely related to the partition function of any other $(j',1-j')$ system,
in particular if $j-j'\in\mathbb{Z}$. Thus, CFT alone is not able to fix $j$.
However, if we take the composite fermion as the basic object, we might expect
that the FQH state involving composite fermions made out of electrons with $p$
attached pairs of flux quanta should correspond to spin $j=\frac{1}{2}(2p+1)$
fields in the CFT description. These should be elementary in the sense that the
spectrum of the CFT does not contain fermionic fields with smaller spin in the
non-twisted sector. Moreover, the twists related to the quasi-particle
excitations should have a minimal charge of $\alpha=1/(2pm+1)$ for an $m$ layer
state, since this is the expected fractional statistics. The fractional charge
is entirely determined by the geometry, i.e., by the number of sheets in the
covering of the complex plane. But the requirement that the composite fermions
shall be the effective elementary particles fixes $j=\frac{1}{2}(2p+1)$ or
$j=\frac{1}{2}(2p+3)$ due to the duality $j\leftrightarrow 1-j$. A very
interesting question is, whether an effective theory of transitions between
different FQH states could yield a mechanism how our CFTs are mapped onto each
other, e.g.~along the lines of \cite{Flohr:1993aa,Flohr:1996aa}.

Finally, we point out that our scheme should be understood as a proposal.
Although we provided a stringent geometrical setting which identified our
choice of CFTs, we cannot connect these CFTs to the full (2$+$1)-dimensional
bulk theory via rigorous first principles. For instance, and in contrast to the
(1$+$1)-dimensional edge theory, there is no mathematical rigorous theorem
which would allow us to invoke some sort of \pstress{Chern-Simons} versus CFT
equivalence. Furthermore, our expressions for the bulk wave functions in terms
of CFT correlators, as all existing proposals for bulk wave functions, should
be understood as trial ones, since exact solutions are not known (this even
applies to the \pstress{Laughlin} wave functions). Comparison with others
obtained from numerical diagonalization of the exact \pstress{Hamiltonian} can
only be made for a small number of electrons and not in the thermodynamic
limit. On the other hand, trial wave functions such as the ones conceived by
\pstress{Laughlin} possess many special features or symmetries, e.g.\
topological order or incompressibility, i.e.\ symmetry under area-preserving
diffeomorphisms. We hope that future research will reveal the physical nature
of such properties such that the connection with CFT is eventually put on
firmer ground and trial wave functions are more thoroughly checked or even
derived from first principles.
%
% Appendix
%
\setcounter{equation}{0}
\renewcommand{\theequation}{\mbox{A\arabic{equation}}}
\section*{Appendix:  Discussion on Unitarity}
It might seem disturbing that the CFTs proposed to describe the FQH bulk regime
are non-unitary. We stress again that these CFTs are not meant to yield the
bulk wave functions from a dynamical principle, nor do they provide an
effective \pstress{Hamiltonian}. Moreover, since the relevant states are
stationary eigenstates of the of a full (2$+$1)-dimensional system, no time
evolution is involved. In this sense, the bulk theory can be reduced to a truly
\pstress{Euclidean} one which is (2$+$0)-dimensional.
The topological nature of the full
(2$+$1)-d system suggests the bulk theory to be at least scale invariant. Thus,
the assumption that bulk wave functions should have a CFT description is
reasonable, but the requirement that these CFTs should be unitary is not
necessary and does not contain any physically relevant information. The bulk
CFT describes purely geometry, namely how the corresponding wave functions can
be understood in terms of vector bundles over \pstress{Riemann} surfaces
\cite{Varnhagen:1994nd}. As we have argued in the main text, the fractional
statistics of the quasi-particle excitations results in a multi-valuedness of
the wave functions, considered as functions over the complex plane. One of the
central features of our approach is to replace this setting by the
geometrically more natural scheme of holomorphic functions over a ramified
covering of the complex plane leading to the non-unitary $(j,1-j)$
\nstress{$b$/$c$-spin} systems.

However, the question of unitarity is not irrelevant. To be consistent, we
should require that our ansatz fits together with the (1$+$1)-d CFTs describing
edge excitations. These describe waves propagating along the one-dimensional
edge of the quantum droplet and hence necessarily have to be
unitary\footnote{This also follows from the strict one-to-one correspondence of
(2$+$1)-dimensional \pstress{Chern-Simons} theories on a manifold $M$ with
unitary (1$+$1)-dimensional CFTs living on the boundary $\partial M$.}.
Consistency requires that the space of states of either CFT, the edge and the
bulk one, should be equivalent. In other terms, both should have the same
partition functions. Fortunately, the \nstress{$b$/$c$-spin} systems have
well-known partition functions which are indeed equivalent to those of certain
${\rm c}=1$ \pstress{Gaussian} models. These latter unitary CFTs are precisely
the candidates for the description of the edge excitations which are most
widely used\footnote{There are some other proposals making use of so-called
minimal ${\cal W}_{1+\infty}$ models or $\widehat{SU}(m)$ \pstress{Kac-Moody}
algebras for $m$-layer states, see for example
\cite{Cappelli:1996pg,Read:1990xx,Wen:uk,Frohlich:wb}.}.

To be more explicit, we consider a spin $(j,1-j)$
\nstress{$b$/$c$-spin} system in some twisted sector with twist $\alpha$.
The full character of this system, including the ghost number, is
defined as
\begin{equation}
  \chi^{(j,\alpha)}(q,z) \equiv {\rm tr}_{{\cal H}^{(\alpha)}}
  q^{L_0^{(j,\alpha)}-\frac{{\rm c}_j}{24}} z^{j_0^{(\alpha)}}\:\: ,
\end{equation}
where we have clearly indicated that the mode expansions of the
\pstress{Virasoro} field
and the ghost current depend on the twist sector. Explicitly computed, these
characters read:
\begin{equation}
  \chi^{(j,\alpha)}(q,z) = q^{\frac{1}{2}(j+\alpha)(j+\alpha+1)+\frac{1}{12}}
  z^{\alpha}\prod_{n=1}^{\infty}
  (1 + zq^{n+(j+\alpha)-1})(1+z^{-1}q^{n-(j+\alpha)})\,.
\end{equation}
It is evident from this formula that the characters (almost) only depend on
$(j+\alpha)$. In particular, we obtain the equivalence:
\begin{equation}
  \chi^{(j,\alpha)}(q,z) = z^{\frac{1}{2}-j}
  \chi^{(\frac{1}{2},\alpha+j-\frac{1}{2})}\:\: .
\end{equation}
Thus, the \pstress{Virasoro} characters (putting $z=1$) of the
\nstress{$b$/$c$-spin} systems are all equivalent to characters of the complex
fermion with ${\rm c}=1$ where the twist sectors $\alpha$ get mapped to others
with $\alpha+j-\frac{1}{2}$. Thus, all sectors which are mapped in this way
keep their statistics, since $j\in\mathbb{Z}+\frac{1}{2}$ and
$\alpha\equiv\alpha+j-\frac{1}{2}$ mod $1$. A more detailed analysis (see,
e.g.~\cite{Guruswamy:1998aa,Kausch:1995aa,Kausch:2000aa,Eholzer:1998aa})
reveals that the partition functions are indeed equivalent. This extends to the
${\rm c}=-2$ spin system describing pairing, which has been pointed out in
\cite{Flohr:1997aa,Cappelli:1998ma}. Therefore, the space of states of
\nstress{$b$/$c$-spin} systems with twists $\alpha=k/m$, $k=0,\ldots,m-1$, is
equivalent to the space of states of a rational ${\rm c}=1$ ($\mathbb{Z}_2$
orbifold) theory with radius of compactification $2R^2=1/m$. The careful
reader should note that this equivalence holds. Although we always consider $m$
copies of our \nstress{$b$/$c$-spin} systems, we work in an \pstress{Abelian}
projection where the charges (or twists) of all copies of the fields are
closely related to each other. Since they are not chosen independently, we only
get one copy of the \pstress{Hilbert} space.
%
%  Section - Acknowledgements
%
\section*{Acknowledgements}
We would like to thank in particular W.~Apel for many helpful discussions,
remarks and encouragement. This work would not have been possible without many
insights on the FQHE he shared with us. We are very grateful to R.\ Haug and U.\
Zeitler for
providing us with up to date experimental facts and details.
Interesting comments and remarks by A.~Altland and L.~Amico
are kindly acknowledged. Finally, we thank A.~Bredthauer and
C.~Sobiella for careful reading of the manuscript. The research of M.F.\ is
supported by the Deutsche Forschungsgemeinschaft, SPP no.\ 1096, Fl 259/2-1.
%
%  The Bibliography
%

\clearpage

\end{document}